\documentclass[lettersize,journal]{IEEEtran}
\usepackage{amsmath,amsfonts}
\usepackage{algorithmic}
\usepackage{algorithm}
\usepackage{array}

\usepackage[caption=false,font=normalsize,labelfont=sf,textfont=sf]{subfig}
\usepackage{listings}
\usepackage{textcomp}
\usepackage{stfloats}
\usepackage{url}
\usepackage{verbatim}
\usepackage{graphicx}
\usepackage{cite}
\usepackage{hyperref}
\usepackage{multirow}
\usepackage{subcaption} % for subfigures
\usepackage{soul}
\usepackage{xcolor}
\sethlcolor{yellow}

\begin{document}

\title{Predicting Human Visual Attention on Words in Source Code}

\author{Chia-Yi Su and Collin McMillan
        % <-this % stops a space
%\thanks{This paper was produced by the IEEE Publication Technology Group. They are in Piscataway, NJ.}% <-this % stops a space
\thanks{Chia-Yi Su and Collin McMillan are with Department of Computer Science and Engineering,
University of Notre Dame, Notre Dame, IN 46556 USA}}

\markboth{Journal of \LaTeX\ Class Files,~Vol.~14, No.~8, August~2021}%
{Shell \MakeLowercase{\textit{et al.}}: A Sample Article Using IEEEtran.cls for IEEE Journals}


\maketitle

\begin{abstract}
This paper presents a computational model to predict human visual attention over words in software source code.  The visual attention of software engineers when reading source code has long been studied as a means to understand human cognitive processes during software engineering tasks.  Predicting this visual attention is important for perfecting user interface design and understanding what information human programmers need.  We propose a model of programmer visual attention in which we design a novel loss function that computes similarity between human attention measured during eye tracking experiments and the internal attention of the artificial neural network.  We evaluate our model by comparing its outputs to actual eye tracking data from three separate datasets. Two are in the Java programming language and one is in the C programming language. Our model outperforms the baseline in software engineering by 64\%, 16\%, and 467\% in each of these studies according to Pearson correlation. We used scanpath prediction as an example to demonstrate that our model is more capable of the task that requires the understanding of human thought process.  Our model achieves a statistically significant improvement over the close baseline in the reading task according to normalized Levenshtein distance and outperforms both Claude and GPT-5 on both reading and writing tasks.
\end{abstract}


%{
%\setlength{\parskip}{0pt}
\section{Introduction}
Human visual attention is a mechanical process that people follow to select particular visual stimuli with their eyes~\cite{carrasco2011visual}.  In this paper, the humans are software engineers and the visual stimuli are words in source code of programs the software engineers read or build.  Visual attention over source code is distinguished from generalized computer vision in that movement and object cues are absent, and it is distinguished from reading text because the eye tends to scan many areas of code with frequent regressions spread over many files with many hundreds of lines of code each~\cite{busjahn2015eye, wallace2025programmer}. The underlying cognitive goal is different: software engineers reading code are building a mental model of software program behavior, versus spacial awareness or story comprehension~\cite{bouraffa2023developers, letovsky1987cognitive, littman1987mental, von2002program}.

Predicting human visual attention in source code involves a forecast of which words in a program's source code will be read most often, reread most often, and/or to what degree and in what order.  Predictions of human attention are useful for determining what information software engineering tools should display~\cite{ko2009finding, zyrianov2020automated}, designing user interfaces~\cite{cheng2012eye}, and developing bio-inspired machine learning models~\cite{bansal2023modeling, zhang2024eyetrans}. Predictions of human attention also holds academic value as a means of studying human cognitive processes~\cite{abid2019using, bednarik2012expertise, karas2024tale, sharafi2015systematic}.  

Typical strategies for predicting visual attention have their origin as either general computer vision research (where the task is variously known as ``gaze tracking'', ``gaze prediction'', etc.~\cite{hu2020dgaze, mazzeo2021deep, zhao2016gaze}) or from models of reading text~\cite{davoudi2015critical, rayner2010models}.  But models from general vision research treat the prediction problem as coordinates in a plane or 3-d space, tending to ignore the semantic information in words or code items.  And reading models assume particular patterns of eye scan which are not found over source code~\cite{busjahn2015eye}.  Lately an alternative has been proposed based on a hypothesis that artificial neural networks natively exhibit similar patterns of attention as humans, but as Eberle~\emph{et al.}~\cite{eberle2022transformer} point out, this hypothesis is not borne out for highly task-specific gaze such as reading source code.

In this paper, we propose a computational model of human visual attention over words in source code.  The input to our model is a snippet of source code, fixation tokens and tokens next to fixation tokens.  The output is a measure of the total gaze time a person is likely to look at each word.  The model predicts the metric \textit{percent total gaze time} (ptgt), which is used as an aggregate metric of attention per word in code in related literature~\cite{bansal2023towards, rodeghero2014improving}.  Our model is novel in two key ways: First, we use eye-tracking datasets to finetune the pretrained language models proposed by~\cite{su2024distilled}. Second, we designed a novel loss function that computes the similarity between human attention (ptgt) and machine attention (self-attention) for finetuning.

We evaluate our model in an experiment comparing the model's predictions to actual eye tracking data.  We source the eye tracking data from studies of two programming languages and two programming tasks to maximize generalizability of our experiment.  Two datasets are studies of Java programmers tasked with reading code to write documentation (the Rodeghero study and Wallace study)~\cite{wallace2025programmer,rodeghero2014improving} and another is a study of C programmers tasked with locating bugs (the Smith study)~\cite{smith2025human}.  We compare our model to the most related and current baseline from software engineering literature~\cite{bansal2023towards} and computer vision literature~\cite{tafasca2024toward}.  We also create an ablation version of our model to study the effect of a novel component of our finetuning procedure. Compared with baseline~\cite{bansal2023towards}, our method achieves 64\%, 16\%, and 467\% improvements in Pearson correlation  on Wallace, Rodeghero, and Smith study.

  %We release all data and implementation details online\footnote{\href{https://anonymous.4open.science/r/human-attention-prediction-0598/README.md}{Anonymous online appendix}}.

We conduct an additional experiment on scanpath prediction to demonstrate the effectiveness of our model on the task that requires the understanding of human thought process. We used the dataset designed for the scanpath prediction task proposed by Bansal~\emph{et al.}~\cite{bansal2023modeling}. We compared our model with the LLMs-based approach proposed by Bansal~\emph{et al.}~\cite{bansal2023modeling} and commercial models. We show that our model outperforms Bansal baseline and both GPT-5 and Claude on this task.

\begin{table}[!t]
    \centering
	{\small
        \setlength{\tabcolsep}{5pt}
		\begin{tabular}{p{4.2cm}p{0.4cm}p{0.4cm}p{0.4cm}p{0.4cm}p{0.4cm}}
        \centering
			&   B         & E   & M & L              \\
            % \citet{curtis1979measuring}& &  	&   &   & x \\
            Rodeghero~\emph{et al.}\cite{rodeghero2014improving}						&   	&  x    & & \\
            Konopka~\emph{et al.}\cite{konopka2015combining}&    	&  x    & & \\
            Fritz~\emph{et al.}\cite{fritz2016leveraging}&   x	&  &   &  \\

            Muller~\emph{et al.}~\cite{muller2016using}&    x	&   & &  \\
            Kevic~\emph{et al.}~\cite{kevic2017eye}&    	& x   & &  \\
            Sharif~\emph{et al.}~\cite{sharif2017eye}&    	&  x   & &  \\
            Abid~\emph{et al.}~\cite{abid2019using}&   	&  x  & & \\
             De Souza~\emph{et al.}\cite{de2020toward}&    x  	&    & &  \\
             Peitek~\emph{et al.}\cite{peitek2021program}&  x  	&     & &  \\
             Sharafi~\emph{et al.}~\cite{sharafi2022eyes}&    	& x     & & \\
             Abbad~\emph{et al.}\cite{abbad2022estimating}& x  	&     & & \\
             Kuang~\emph{et al.}~\cite{kuang2023applying}&    	&     &x & \\
             Sun~\emph{et al.}~\cite{sun2023program}&     	&  & x  & \\
             Yoshioka~\emph{et al.}~\cite{yoshioka2025eye2vec}&    &	  &  x & \\
            Bansal~\emph{et al.}~\cite{bansal2023towards}&     & x	  & x  &  \\
            Bansal~\emph{et al.}~\cite{bansal2024modeling}  &	 & x   & & x  \\
            Paltenghi~\emph{et al.}~\cite{paltenghi2024follow}&    	 & &  x  &  \\
            Grandel~\emph{et al.}~\cite{grandel2025comcat} &  	  &   & & x \\
            Zhang~\emph{et al.}~\cite{zhang2025eyemulator} &  	  &  &  & x \\
            %\citet{gao2025nrevisit}&    x	&  &   &  \\
            \emph{this paper}& & x  & & x \\
		\end{tabular}
	}
	%\vspace{0.2cm}
	\caption{Snapshot of the key papers related to programmer's visual attention prediction. $B$ means biometric-based approaches.  $E$ means eyetracking-based approach. $M$ means classic machine learning-based approach. $L$ means LLM-based approach.}
	\label{tab:phrelated}
    %\vspace{-2mm}
\end{table}

\section{Background \& Related Work}
In this section, we discuss the research related to  predicting human attention during software engineering tasks and scanpath prediction. We summarize related papers on predicting human attention during software engineering tasks in Table~\ref{tab:phrelated}. 

\subsection{Biometric-based Approaches}
Biometric-based approaches predict programmer attention based on measurements of body activity~\cite{fritz2016leveraging}.  These approaches tend to target coarse-grained signals of attention. The goal is to predict whether a person is, e.g., encountering a stressful workload~\cite{calvo2010affect}.  Biometric-based sensing of engagement and attention level has long been studied in other domains~\cite{monkaresi2016automated}, and has been applied to programmers~\cite{muller2016using}. Likewise, Fritz~\emph{et al.}~\cite{fritz2016leveraging} trained a Naive Bayes classifier to predict the difficulty of programming tasks and programmer resilience to intellectual challenges.  

\subsection{Eyetracking-based Approaches}
Eye-tracking is a technology that measures the location of a person's gaze over time, along with other eye data such as pupil dilation~\cite{carter2020best}.  Eye-tracking has been used for decades to study human attention during many types of tasks including software engineering -- the community is mature enough to have guides describing best practices for eye tracking in software engineering~\cite{burch2025eye, sharafi2020practical} and surveys chronicling its history~\cite{grabinger2024eye, obaidellah2018survey, sharafi2015systematic}.  However, \emph{predicting} human programmer visual attention is much less common.  An early example uses eye tracking data as a guide for crafting weight modifications by hand to an information retrieval system that generates source code documentation~\cite{rodeghero2014improving}.  A different group of researchers aimed to predict programmer navigation through code by identifying influential visual features during software engineering tasks such as bug localization and traceability link recovery~\cite{sharif2017eye, sharafi2022eyes}.  Nuances of behavioral signals have been studied as well, such during code changes, cross-language environments, and scanpaths~\cite{abid2019using, bansal2023modeling, kevic2017eye, konopka2015combining}.  Yet by far the most similar work to this paper predicts the degree of human visual attention using a graph neural network (GNN).  The graph input to the GNN is an abstract syntax tree (AST) of a method written in the Java programming language, and the output is a measure of the amount of gaze time spent on each word in the method~\cite{bansal2023towards}.  We use this GNN approach as a baseline in our evaluation later in this paper.

\subsection{Computer Vision-based Approaches}
General computer vision-based approaches formulate these problems as gaze prediction~\cite{tafasca2024toward} and the prediction of neural response  in visual systems~\cite{wang2025cognitive,saha2024modeling}.  The input of these approaches is in the 2-d or 3-d space. The goal is to predict the target area or compare the models with human brains. However, these approaches tend to ignore the semantic information in vocabularies, e.g., source code~\cite{xu2014predicting}. In this paper, we propose an alternate computational model for source code.

\subsection{ML and LLM-based Approaches}
Recently, different approaches to predict programmer attention based on classical machine learning (ML) techniques have given way to approaches using large language models (LLMs).  Examples in this vein include efforts to predict the next fixation token during code navigation using recurrent neural models (RNNs)~\cite{sun2023program} and an approach that uses a GPT2-like language models to predict the scanpath of programmers during code documentation~\cite{bansal2023modeling}.  Another strategy is to modify the structure of the attention mechanism of artificial neural networks based on experimental observations of programmers~\cite{zhang2024eyetrans, zhang2025eyemulator}.  These approaches tend to be highly novel but have niche applications such as attention to function call graphs during code summarization~\cite{McLoughlin2025VisualAttentionFCG} or prediction of attention levels for the purpose of studying the similarity of machine and human attention~\cite{li2024machines}.

\subsection{Summary}
Table~\ref{tab:phrelated} shows a trend from biometric, through hand-crafted eye-tracking-based, to data-driven ML and LLM-based approaches most recently.  While eye-tracking data itself has been used consistently, these have been used as training inputs to ML and LLM techniques rather than as a basis for hand-crafted features. The gap in the literature this paper fills is predicting attention at a fine level of granularity (specific words in source code) using an LLM-based computational model trained with eye-tracking data.

\subsection{Scanpath prediction}

Scanpath is a  is a task of predicting sequence of fixations made by a person while observing a visual stimulus~\cite{bansal2023modeling, deng2023eyettention, reichle1998toward, engbert2005swift, nilsson2009learning, nilsson2011entropy}. Before neural models, Reichle~\emph{et al.}~\cite{reichle1998toward} proposed E-Z Reader that models a eye-movement based on the reading pattern for different types of word and passages.  Alternatively, Engbert~\emph{et al.}~\cite{engbert2005swift} implemented a mathematical models to model the eye movement in the general reading comprehension. More recently, neural-based approaches for scanpath prediction have been proposed. For example, Deng~\emph{et al.}~\cite{deng2023eyettention} proposed a LSTM-based approach that consider different reading pattern, e.g., non-left-to-right reading pattern. However, these approaches focus on reading comprehension instead of programming tasks. Bansal~\emph{et al.}~\cite{bansal2023modeling} first proposed a GPT2-like language model to predict the scanpath of human programmers while doing source code summarization. Therefore, we use this as our baseline models.

\section{Computational Model}\label{sec:computation_model}

Our computational model is a Transformer neural architecture that we train with a custom loss function.  We finetune a pretrained Transformer using actual eye-tracking data over source code.  Our loss function for finetuning takes the machine attention vectors in the last layer of the Transformer model, aggregates those vectors for fixation, then computes a similarity between the machine's attention and a reference human visual attention calculated for the fixation tokens in the input.  This section discusses this custom loss function, the ingredients required to work, and the model's prediction target.

\subsection{Prediction Target}

Our prediction target is \emph{percent total gaze time (ptgt)}.  The \emph{ptgt} metric was proposed by Rodeghero~\emph{et al.}~\cite{rodeghero2014improving} to quantify the ``importance'' of a word in source code to a person reading that code. Formally, \emph{ptgt} is:
\vspace{-2mm}
\[
\mathrm{ptgt}(r) \;=\; \frac{G_r}{G_{\text{total}}}.
\]

Where $G_r$ denotes the total gaze time (in milliseconds) spent on a region of interest (ROI) $r$ (where $r$ is usually a word in source code), and $G_{\text{total}}$ denotes the total gaze time spent viewing the entire code artifact (e.g., all the words in a code subroutine).  Percent total gaze time quantifies the proportion of a programmer’s visual attention allocated to a specific code region relative to all viewed code. By normalizing gaze time by the total viewing duration, \emph{ptgt} enables comparisons across different regions, methods, and participants independent of absolute reading time or size of the ROI.

\subsection{Ingredients}

The two key ingredients to our computational model are:

\begin{enumerate}
    \item Pretrained Transformer-based Language Model.  Almost all current large language models are based on the Transformer architecture in one form or another, with the main difference among models being the size of the model and the data on which it is trained.  While it is possible to train our computational model from scratch, the large amount of data required to train language models and the relatively small amount of available eye-tracking data leads us to finetune a pretrained model.  The Methods section describes the pretrained language model in our experiments.

    \item Eye-tracking Data over Source Code.  We require samples of actual human visual attention to calculate \textit{ptgt} and finetune the pretrained model.  We obtain these samples from experiments of human programmers doing software engineering tasks at a computer equipped with an eye-tracking device.  The samples include source code and human gaze data.  We used leave-one-out validation, i.e., we holdout one participant at a time and averaged results. We describes our  datasets in Section~\ref{sec:datasets}.
    
\end{enumerate}

The input sequence is composed of words from a section of source code, fixation tokens, and tokens next to the fixation tokens code. We tokenized input sequence with the GPT2 tokenizer as in~\cite{su2024distilled}:

\vspace{-2.5mm}
\[
\mathbf{S} = (x_1, \dots, x_{T_x},\; y_1, \dots, y_{T_y},\; z_1, \dots, z_{T_z})
\]
Where $x_{t_x}$ is the source code tokens, $y_{t_y}$ is the fixation tokens, and $z_{t_z}$ is the tokens next to the fixation tokens. The model $\mathcal{M}^{\text{pre}}$ is pretrained using the typical categorical cross-entropy (CCE) loss denoted $\mathcal{L}_{\text{CCE}}$.

\subsection{Custom Loss Function}

To create our computational model, we finetune $\mathcal{M}^{\text{pre}}$ to create $\mathcal{M}^{\text{ft}}$ using a our custom loss function, $\mathcal{L}$.  During finetuning, we calculate loss in three steps:

\subsubsection{Attention Readout}
Our attention readout mechanism extracts the model's attention to each fixation token in the input sequence. In a Transformer model, $Q$ and $K$ are ``attention'' matrices used to scale the value matrix $V$, which represents the input tokens.  A typical LLM will have many Transformer layers -- what we do is extract $Q$ and $K$ from the final Transformer layer (note, Wang~\emph{et al.}~\cite{wang2024probing} found that the attention heads in the final layer are most like human attention). Formally, in the final transformer layer with $H$ heads and input sequence length $T$, the model computes query, key, and value matrices for each attention head as:

%Attention readout is important because it provides an overview of how the model does internally compared with humans~\citep{saha2025modeling}.

%Our attention readout mechanism extracts the model's attention to each fixation token in the input sequence. In a nutshell, we extract attention values from the final Transformer layer because~\citet{wang2024probing} found that the attention heads in the final layer are most like human attention. Formally, in the final transformer layer with $H$ heads and input sequence length $T$, the model computes query, key, and value matrices for each attention head as:
\vspace{-5mm}
\[
Q^{h},K^{h} \in \mathbb{R}^{ T \times d_v},\; V^{h} \in \mathbb{R}^{ T \times d_v}, \quad \text{where } d_v=d_{\mathrm{model}}/H
.
\]
Each scaled dot-product attention is computed as:
\vspace{-2mm}
\[
A^{h} = \operatorname{softmax}(\frac{Q^{h} {K^{h}}^\top}{\sqrt{d_k}}),
\quad
A^{h} \in \mathbb{R}^{T \times T}.
\]
We obtain our attention readout by averaging the attention scores from attention heads 10 to 16. Wang~\emph{et al.}~\cite{wang2024probing} observed that attention heads in deeper levels are closer to human attention. Since our task is to predict human  attention, we used  attention heads from 10 to 16. We formally define our attention readout as:
\vspace{-2.5mm}
\[
\mathbf{\bar{A}}
=
\frac{1}{H-10} 
\sum_{h=10}^{H} 
A^{(h)},
\]
Each row of $\bar{A}$ represents a probability distribution over key tokens attended by a given query token.

\subsubsection{Model-Derived Attention Score}

Next, we define an attention score by averaging attention probabilities from a subset of query tokens and key tokens in $\bar{A}$. Formally, we define the attention score as:
\vspace{-1mm}
\[
\hat{\mathbf{a}}
=
\frac{1}{|\mathcal{Q}'|\,|\mathcal{K}'|} 
\sum_{t_z \in \mathcal{Q}'} 
\sum_{t_y \in \mathcal{K}'} 
\bar{A}_{t_z,t_y}, 
\quad 
\mathcal{Q}' \subseteq \mathcal{Q}, 
\quad 
\mathcal{K}' \subseteq \mathcal{K}.
\]
Where $\mathcal{Q}'$ is a set of tokens next to the fixation tokens and $\mathcal{K}'$  is a set of the fixation tokens in the input. This scalar represents the probability mass that the fixation tokens are attended to by its next token. %We show one example in Appendix.%The goal is to train the model that 

\subsubsection{Human Attention Supervision}

Finally, we compute the similarity of the predicted attention $\mathbf{\hat{a}}$ to the reference human attention $\mathbf{a}$. Reference human attention $\mathbf{a}$ is \textit{ptgt} for each token (see the Metrics section under Methods for a complete definition of $\mathbf{a}$). Then, we combine that similarity with the standard CCE loss.  To compute the similarity of $\mathbf{\hat{a}}$ to $\mathbf{a}$, we use mean squared error (MSE) and Pearson correlation.  Our rationale is that MSE will enforce absolute accuracy of the predicted fixation time, while Pearson correlation will enforce the relative attention structure (high vs. low attention tokens).  The combination encourages the model to predict correct magnitude and to preserve human-like attention distributions across tokens.  Formally, we use MSE as:
\vspace{-2mm}
\[
\mathcal{L}_{\text{MSE}}
=
\frac{1}{m}\sum_{i=1}^{m}(\hat{a}_i - a_i)^2,
\]
Where \(m\) is the batch size. 
\begin{comment}
We center both $\hat{a}$ and $a$ as:
\vspace{-2mm}
\[
\tilde{\hat{a}}_i = \hat{a}_i - \frac{1}{m}\sum_{k=1}^m \hat{a}_k,
\quad
\tilde{a}_i = a_i - \frac{1}{m}\sum_{k=1}^m a_k.
\]

We compute Pearson correlation as:
\vspace{-2mm}
\[
\rho =
\frac{\sum_{i=1}^m \tilde{\hat{a}}_i \tilde{a}_i}
{\sqrt{\sum_{i=1}^m \tilde{\hat{a}}_i^2 + \varepsilon}
 \sqrt{\sum_{i=1}^m \tilde{a}_i^2 + \varepsilon}}.
\]
\end{comment}
We denote $\rho$ as the Pearson correlation. We normalize and convert $\rho$ to a loss as:
\vspace{-2mm}
\[
\mathcal{L}_{\text{corr}} = 1 - \tfrac{1}{2}(\rho + 1).
\]

To prevent the auxiliary attention losses from being dominated by the language modeling loss, we dynamically rescale them to match the magnitude of the cross-entropy loss.  The final training objective is:
\vspace{-2mm}
\[
\mathcal{L}
=
\mathcal{L}_{\text{CCE}}
+
\lambda_{\text{MSE}}\,\alpha_{\text{MSE}}\,\mathcal{L}_{\text{MSE}}
+
\lambda_{\text{corr}}\,\alpha_{\text{corr}}\,\mathcal{L}_{\text{corr}},
\]
where \(\alpha_{\text{MSE}}\) and \(\alpha_{\text{corr}}\) are scaling factors based on the ratio between the cross-entropy loss and the corresponding attention loss, and \(\lambda_{\cdot}\) are fixed dominance coefficients.  Table~\ref{tab:hyperparams} gives the values for all the training hyperparameters in our evaluation. We used these hyperparameters based on our experiments.

\subsection{Finetuning}

We follow conventional finetuning procedures for Transformer-based language models of source code established in many years of related work~\cite{naveed2025comprehensive, su2023language, xu2022systematic}.  Note above that our loss function uses $\mathcal{L}_{\text{CCE}}$ as is conventional -- the model's goal is still to learn to generate samples in an autoregressive fashion like almost all current language models.  The model will see source code samples from the eye-tracking experiments as though they are normal samples during normal training.  However, the effect of our loss function is to increase or decrease the penalty the model receives based on how similar its own internal attention is to human visual attention observed experimentally.  This process yields a model $\mathcal{M}^{\text{ft}}$ which is operationally identical to any other autoregressive language model but which has attention scaled to be more like a human programmer.  Note, however, that we can extract $\mathbf{\hat{a}}$ as above to obtain attention predictions from the model after finetuning.

\begin{table}[b]
\centering
\small
\vspace{-2mm}
\caption{Computational model training hyperparameters for three datasets.}
\vspace{-2mm}
\label{tab:hyperparams}

{\setlength{\tabcolsep}{1.5pt}
\renewcommand{\arraystretch}{1.2}
\begin{tabular}{@{}lcccc@{}}
\hline
Hyperparameter & Symbol & Wallace & Smith & Rodeghero \\
\hline
Training epochs      & $N_{\text{epochs}}$ & 9 & 5 & 7 \\
Batch size           & $B$                 & 8 & 8 & 4 \\
Learning rate        & $\eta$              & 7e-6 & 3e-6 & 1e-5 \\
Embed. size          & $E$                 & 1,024 & 1,024 & 1,024 \\
Number of heads      & $H$                 & 16 & 16 & 16 \\
Number of layers     & $L$                 & 24 & 24 & 24 \\
Gradient accumulation& $k$                 & 128 & 128 & 128 \\
\hline
Loss weights         & $\lambda_{\text{MSE}}, \lambda_{\text{corr}}$ & 5 & 5 & 5 \\
%Scaling factors      & $\alpha_{\text{MSE}}, \alpha_{\text{corr}}$   & \multicolumn{3}{c}{$\text{clip}\Big(\frac{\text{CCE}}{\text{MSE/Corr Loss}}, 1e-3, 1e6\Big)$} \\
\hline
\end{tabular}}
\vspace{2mm}
\end{table}

\section{Aggregated Attention Prediction}

This section describes our research methods for studying machine and human attention using the computational model in Section~\ref{sec:computation_model}.  Our research objective is 1) demonstrating the degree to which our model mimics the attention people pay to those words by comparing attention on words in source code with humans and our computational model 2) comparing our model to the most-similar computational baselines. To this end, we ask the following Research Questions (RQs):
\begin{description}
    %\item[~~RQ1] What is the performance of open-source LLMs on code development tasks when finetuned for those tasks under recommended conditions?
    \item[~~RQ1] What is the performance of our computational models compared with the baseline and human reference?
    \item[~~RQ2] What is the performance of our computational models with the Pearson correlation loss compared with the one without Pearson correlation loss?
   
\end{description}

The rationale behind RQ1 is that we use the attention scores found to be closer to human attention to finetune the models. However, it is possible that the models do not learn the meaningful information during finetuning, so the baseline could still outperform our model or the results could still be far from human reference. We compare our model with the baseline from software engineering and computer vision.

The rational behind RQ2 is that it is possible that the Pearson correlation loss does not help the model to learn the meaningful information. To this end, we conduct an ablation study to compare the model with the Pearson correlation loss and the one without Pearson correlation loss. 

\subsection{Datasets}
\label{sec:datasets}

We use three datasets in our experiment.  The first dataset (the \textbf{Smith study}) involves 21 programmers who examined up to eight memory bugs in three different C programs on a machine equipped with a Tobii Pro Fusion eye-tracking device~\cite{smith2025human}. The second dataset (the \textbf{Wallace study}) involves 10 programmers who read 40 subroutines in the Java programming language in five different sessions~\cite{wallace2025programmer}. The task is to read Java source code and write an English sentence that describes the purpose of that code on a machine equipped with a Tobii Pro Fusion eye-tracker.  The third dataset (the \textbf{Rodeghero study}) involves 10 professional programmers who were hired to read Java source code and write an English sentence that describes the internal of that function on a machine equipped with a Tobii TX300 eye-tracking device~\cite{rodeghero2014improving}. All studies are independent. We chose these three different studies to maximize the generalizability of this study. 

\subsection{Computational Model Settings}

We use the pretrained language model \texttt{jam} as $\mathcal{M}^{\text{pre}}$ in this study.  The \texttt{jam} model is a GPT-2 style language model that has around 350m parameters~\cite{su2023language}.  We select this model for two reasons.  First, it includes Java and C versions each with a tightly-regulated training set intended for controlled experiments on software engineering datasets.  Second, subsequent experiments demonstrated that the 350m parameter model is a ``goldilocks'' size in that it is large enough to be competitive with large commercial models on select software engineering tasks, and yet small enough to be trainable on commercial hardware~\cite{su2024distilled}, which is important to ensure reproducibility of results and control of experimental variables. Note that we finetune Rodeghero dataset with the model that has been finetuned with Wallace dataset because its size is smaller.  %We train our model with the hyperparameters in Table~\ref{tab:hyperparams}.%Trainability on commercial hardware is important to ensure reproducibility of results and control of experimental variables (i.e., risks from unknown changes in online platforms or artifacts of additional training steps introduced by parameterization such as LoRA).  %We train our model with the hyperparameters in Table~\ref{tab:hyperparams}.

%\subsection{Weight Freezing}
We freeze parts of the weights to finetune our models with the proposed loss function $\mathcal{L}$. The purpose is to increase the batch size because Pearson correlation is less effective when the batch size is small~\cite{wong20223d}. We finetune layer 14 to 24 because finetuning with only the final layer may not be large enough for models to learn meaningful information and Wang~\emph{et al.}~\cite{wang2024probing} found that these layers better align with human attention compared with other layers.

\subsection{Comparisons with Human Reference}

We compare the human participants in one study to other participants in the same study, e.g. one participant in the Smith study to others in the Smith study.  The purpose of this comparison is to establish a baseline for expected similarity of the computational model to human participants.  We would not expect the computational model's attention similarity to substantially exceed the attention similarity that humans have to each other -- an excellent benchmark would have similarity near to that which humans have to one another.

%Therefore, we calculate the metrics described below for each person in the study.  We report an ``aggregate'' human attention similarity for each study.  We calculate this aggregate by averaging the similarity of each participant to all others for all stimuli in the each study.  By similarity we mean each of the metrics we describe below.  We also report ``individualized'' similarity scores for each participant to all other participants, but we calculate this individualized score only for stimuli that all participants saw.  In the C study, eight software bugs were subjects, of which two were seen by all participants (only some of the participants saw each of the other six).  In the Java study, forty subroutines were subjects, of which five were seen by all participants.  These aggregate and individualized scores provide a coarse- and fine-grained view of attention and are in line with metrics reported by related work~\citep{bansal2023towards}, though this related work only evaluated a much smaller and older dataset~\citep{rodeghero2014improving}.

\subsection{Comparison with Baselines}

We use two previous machine models of attention for comparison, plus one variant of our model as an ablation test.  One is the software engineering model proposed by Bansal~\emph{et al.}~\cite{bansal2023towards}.  A second comparison is the most-similar model from the computer vision research community~\cite{tafasca2024toward}.  Almost all general-purpose computer vision models are not suitable for the task we present in this paper because they use 2-d images as input and output coordinates in those images -- they rely on color, shape, and other real-world visual cues not in source code.  However, the model by Tafasca~\emph{et al.}~\cite{tafasca2024toward} includes a context encoder for semantic information akin to that in words in source code.  We create a faithful re-implementation of both approaches using source code words as input and \emph{ptgt} as output.  These models are useful as comparison points for our model while also showing the trainability of different artificial neural models to mimic human attention.

We also create a variant of our model which is identical except that the loss function does not include a correlation component as an ablation study.  We denote this variant \texttt{ours-nocorr} when reporting results.  Formally speaking, the variant's loss function is:
\vspace{-2mm}
\[
\mathcal{L}
=
\mathcal{L}_{\text{CCE}}
+
\lambda_{\text{MSE}}\,\alpha_{\text{MSE}}\,\mathcal{L}_{\text{MSE}}
\]

\subsection{Metrics}

We use four different measures of correlation as similarity metrics.  We calculate each metric the same way, as an aggregate comparison of \emph{ptgt} predicted by the computational model for each word, to the actual \emph{ptgt} for each word observed experimentally.  However, a mismatch between machine and human attention is that the model outputs one \emph{ptgt} score for each word, but each study has multiple people as participants and these participants do not look at all the same words, and often the same person reads the same word multiple times.  We solve this mismatch by aggregating the \emph{ptgt} for each word for each participant in each session.  The correlation is calculated over pairs, which have the average of all \emph{ptgt} values for a word from a single experimental session involving one participant, and the predicted \emph{ptgt} value for that word.

We have \emph{canonical} word tokens and \emph{virtual} word tokens for our comparison as follows:

\subsubsection{Canonical Word Tokens} We define a \emph{canonical word token} as the pair $(f,t_{src})$.  The set of all canonical word tokens is:
\vspace{-2mm}
\[
\mathcal{W} = \{(f,t_{src}) : f \in \mathcal{F},\; t_{src} = 1,\dots,L_f\}.
\]
Where $\mathcal{F} = \{1,\dots,|\mathcal{F}|\}$ denotes the set of functions (subroutines), and function $f \in \mathcal{F}$ contains $L_f$ word tokens indexed by $t \in \{1,\dots,L_f\}$. Also, we define $W'$ as the subset of $W$, which represents a set of canonical fixation tokens in a function $f$.

We extract a single attention score for each canonical fixation word token from the model. These predictions are invariant across participants and sessions:
\vspace{-2mm}
\[
\hat{a}_{f,t_{src}} \in \mathbb{R}
\qquad
\forall (f,t_{src}) \in \mathcal{W}',
\quad
\mathcal{W}' \subseteq \mathcal{W}.
\]

\subsubsection{Human Study Sessions} From a human study, we have $\mathcal{S} = \{1,\dots,|\mathcal{S}|\}$ experimental sessions.  Each session $s \in \mathcal{S}$ could have multiple participants. Thus, the set of canonical fixation word tokens observed in a session $s$ may be denoted:
\vspace{-2mm}
\[
\mathcal{W}'_s = \{ (f, t_{src}) \in \mathcal{W}' \}.
\]

For each observed canonical fixation word token in a session $(f,t_{src}) \in \mathcal{W}'_s$, we compute \emph{ptgt}, denoted:
\vspace{-2mm}
\[
a_{s,f,t_{src}} \in \mathbb{R}
\]

\subsubsection{Virtual Word Tokens} We define a \emph{virtual word token} as the triple $(s,f,t_{src})$, representing an observation of canonical token $(f,t_{src})$ in a session $s$. The set of all virtual tokens is:
\vspace{-2mm}
\[
\mathcal{V} = \{(s,f,t_{src}) : s \in \mathcal{S},\; (f,t_{src}) \in \mathcal{W}_s\}.
\]

If a canonical word token $(f,t_{src})$ appears in multiple sessions, then it gives rise to multiple virtual tokens, each with its own human attention measurement but sharing the same model prediction. We define a subset of virtual word tokens that represents the set of virtual fixation word tokens as ${W}'_s$.

\subsubsection{Vector Construction for Correlation} We build two vectors, $\mathbf{a}$ and $\hat{\mathbf{a}}$, as an input to a correlation metric.  We let $N = |\mathcal{V}|$ and construct the two vectors of length $N$:
\vspace{-2mm}
\begin{align*}
\mathbf{a} &= \big( a_{s_i,f_i,t_{\text{src}_i}} \big)_{i=1}^{N}, \\
\hat{\mathbf{a}} &= \big( \hat{a}_{f_i,t_{\text{src}_i}} \big)_{i=1}^{N}.
\end{align*}

$\hat{\mathbf{a}}$ contains repeated values whenever the same canonical word token appears in multiple sessions.

\subsubsection{Correlation Computation}

Finally, we compute the correlation between model predictions and human attention over all virtual tokens:
\vspace{-2mm}
\[
r = \mathrm{Corr}(\hat{\mathbf{a}}, \mathbf{a}),
\]
Where $\mathrm{Corr}$ is one of four correlation metrics.

%Where $\mathcal{S} = \{1,\dots,|\mathcal{S}|\}$ denotes the set of experimental sessions,  each session $s \in \mathcal{S}$ correspond to a single participant, and $I_s(f) \in \{0,1\}$ indicate whether function $f$ appears in session $s$.

\subsubsection{Calculation for Human Comparisons}
For human comparisons, we follow the same procedure as above, except that the model prediction is replaced by the attention of a held-out participant. Let $p^\ast$ denote the held-out participant. For each virtual fixation word token $(s,f,t_{src})$ associated with $p^\ast$, we define $a$ and $\hat{a}$ as:
\vspace{-2mm}
\[
\hat{a}_{f,t_{src}} \;=\; a_{p^\ast,f,t_{src}},
\qquad
a_{s,f,t_{src}} \;=\; a_{s,f,t_{src}} - a_{p^\ast,f,t_{src}},
\]
Where $a_{p^\ast,f,t_{src}}$ is the attention score of participant $p^\ast$ for word token $(f,t_{src})$, and $a_{s,f,t_{src}}$ ranges over all other participants’ observations of the same token. We then compute the correlation between $\hat{a}$ and $a$ as before.

\subsubsection{Correlation Functions}  We use the following correlation metrics for the function $\mathrm{Corr}$.  We provide our rationale for each as follows:

\begin{itemize}
    \item \textbf{Pearson correlation} measures linear association between model and human attention values.  It captures whether the model assigns proportionally higher attention to words that humans attend to more, assuming a linear relationship to absolute magnitudes.

    \item \textbf{Spearman rank correlation} evaluates monotonic agreement between model and human attention rankings. It captures whether the model ranks words correctly from most to least attention.%, with less emphasis to magnitude.

    \item \textbf{Kendall’s $\tau$} measures ordinal agreement based on pairwise concordance between rankings. It provides a more conservative estimate of rank agreement than Spearman and is particularly robust to ties and small-sample effects common in token-level attention data.

    \item \textbf{Cosine similarity} measures directional alignment between attention vectors, independent of their overall magnitude. It captures whether the model and humans distribute attention across words in a similar pattern, even when absolute attention values differ.
\end{itemize}

All things considered, these metrics assess agreement in attention \emph{magnitude} (Pearson), \emph{ordering} (Spearman, Kendall), and \emph{distributional shape} (Cosine) between machine-predicted and human attention.

%\subsection{Threats to Validity}
% looks like most CCN papers put this in the Discussion section as Limitations

\section{experiment results}
This section discuss our experiment results.

\subsection{RQ1: Comparison with Human Reference and Baselines}
Table~\ref{tab:shape-metrics} showcases our results.  Our model reaches a higher level of similarity to the human reference attention in all three datasets.  However, we find that all models achieve a degree of similarity to the reference, implying that human visual attention patterns over source code are at least somewhat learnable by a range of artificial neural models.  We find that human visual attention is partially consistent according to our measurement metrics, and machine models even exceed similarity to the aggregated reference than some humans do. %We present these results in this section and discuss higher-level interpretations in the next section.

\vspace{1mm}
\textbf{Human Reference Similarity}
We report aggregate human reference similarity for each study in Table~\ref{tab:shape-metrics} for a comparison point, but with the caveat that differences among individual participants can be large.  Consider the Table~\ref{tab:human-reference-sim} which shows Pearson correlation for each participant to all other participants for the Java methods in the Wallace study. Two points stand out.  First, no participant shows uniformly high agreement across all methods.  However, second, correlation tends to be in the neighborhood of positive 0.40 throughout.  These points hint at a signal of human attention patterns hidden among noisy individual, task, and artifact-dependent factors.  Thus, the human reference we report should be interpreted as a ``ballpark'' comparison point for our model rather than as an idealized notion of perfect agreement.

\begin{table}[h]
\centering
\caption{Pearson correlation for each participant to all other participants for the Java methods in the Wallace study. Some samples were removed during data validation reported by Wallace~\emph{et al.}~\cite{wallace2025programmer}. Participant indices start at 3 to be consistent with original study data.}
\label{tab:human-reference-sim}
\vspace{2mm}
{\normalsize
\setlength{\tabcolsep}{3.5pt}
\begin{tabular}{c lllllllllll}
 & & \multicolumn{10}{c}{participant} \\
 &  & 3 & 4 & 5 & 6 & 7 & 8 & 9 & 10 & 11 & 12 \\
\cline{3-12}
\multirow{5}{*}{\rotatebox{90}{method}} 
& A & \multicolumn{1}{|l}{.24} & .30 & .08 & - & .34 & .38 & .31 & .29 & .22 & .25 \\
& B & \multicolumn{1}{|l}{.39} & .30 & .21 & .35 & .21 & .30 & .48 & .39 & .44 & .44 \\
& C & \multicolumn{1}{|l}{.54} & .34 & .35 & .41 & .41 & .47 & .52 & .45 & .57 & - \\
& D & \multicolumn{1}{|l}{.48} & .48 & .48 & .47 & - & .48 & .23 & .28 & .25 & .47 \\
& E & \multicolumn{1}{|l}{.30} & .42 & .30 & .44 & - & .41 & .43 & .42 & .35 & -
\end{tabular}
}
\vspace{3mm}
\end{table}

\begin{table*}[t]
\centering
%\small
\caption{Experimental results across different approaches and datasets.}
\label{tab:shape-metrics}
\vspace{-2mm}

\begin{tabular}{l l c c c c}
\textbf{Dataset} & \textbf{Model} & \textbf{Pearson $r$} & \textbf{Spearman $\rho$} & \textbf{Kendall $\tau$} & \textbf{Cosine Sim.} \\
\hline
% Java
\multirow{5}{*}{Wallace~\emph{et al.}~\cite{wallace2025programmer}} &
Human Reference            & 0.3829 & 0.3807 & 0.2703 & 0.6416 \\
& Bansal~\emph{et al.}~\cite{bansal2023towards}       & 0.2738 & 0.3736 & 0.2524 & 0.5003 \\
& Tafasca~\emph{et al.}~\cite{tafasca2024toward}       & 0.2066 & 0.2372 & 0.1602 & 0.5146 \\
& ours-no corr                  & 0.3313 & 0.4559 & 0.3128 & 0.5276 \\
& ours                          & 0.4496 & 0.4728 & 0.3271 & 0.6625 \\
\hline
% C
\multirow{5}{*}{Smith~\emph{et al.}~\cite{smith2025human}} &
Human Reference              & 0.2134 & 0.2923 & 0.1999 & 0.4231 \\
& Bansal~\emph{et al.}~\cite{bansal2023towards}     & 0.0438 & 0.1003 & 0.0672 & 0.2906 \\
& Tafasca~\emph{et al.}~\cite{tafasca2024toward}     & 0.0428 & 0.0898 & 0.0597 & 0.3301 \\
& ours-no corr                  & 0.1331 & 0.2668 & 0.1818 & 0.3348 \\
& ours                          & 0.2484 & 0.3381 & 0.2310 & 0.3829 \\
\hline
% Java (2013)
\multirow{5}{*}{Rodeghero~\emph{et al.}~\cite{rodeghero2014improving}} &
Human Reference               & 0.5252 & 0.4583 & 0.3216 & 0.7307 \\
& Bansal~\emph{et al.}~\cite{bansal2023towards}     & 0.4138 & 0.2297 & 0.1548 & 0.7018 \\
& Tafasca~\emph{et al.}~\cite{tafasca2024toward}     & 0.4209 & 0.2985 & 0.2031 & 0.6981 \\
& ours-no corr                  & 0.4317 & 0.4703 & 0.3263 & 0.7076 \\
& ours                          & 0.4828 & 0.4864 & 0.3394 & 0.7218 \\
\hline
\end{tabular}
\vspace{-2mm}
\end{table*}

\vspace{1mm}

\textbf{Model Similarity to Human References} We observe that our model has the highest similarity to a human reference in all three of the datasets and all four metrics.  For example, Pearson correlation is 0.45 in the Wallace study versus 0.27 for one baseline and 0.21 for another. One possible explanation is our readout mechanism for extracting attention from the final transformer layer. Indeed, Saha~\emph{et al.}~\cite{saha2025modeling} conclude that when modeling the visual system with artificial neural networks, the ``choice of readout mechanism significantly impacts prediction accuracy.''  We extract attention directly from the component of the model intended to mimic and leave the model's original autoregressive design intact.

At times the machine models \emph{exceed} the human reference according to the correlations we calculated.  What this means is that, for certain functions and words, the model's predicted attention aligns more consistently with the aggregate human attention pattern than any single individual participant aligns with the rest of the group.  This does not imply that the model is ``more human'' than the humans.  It reflects that the model estimates human attention overall, whereas individual human measurements are subject to personal variability.  Note that at all times cosine similarity is the highest among humans, implying that people distribute attention across words in more similar patterns to each other than a model.  This observation is consistent with findings in other cognitive studies, where averaging across observers reveals robust attentional structure that may be obscured at the level of single-subject measurements~\cite{hasson2004intersubject, hasson2010reliability, nastase2019measuring}.

\textbf{Dataset Differences} Similarity measures are overall lower in the Smith dataset than in the Wallace or Rodeghero datasets.  We note that in some cases model performance is quite low over the Smith dataset, as the baselines appear to struggle to have a positive correlation at all.  Even the human reference attention similarity is relatively low.  One likely explanation is the nature of the C programming language itself.  Functions in C tend to be shorter than in Java, and the words used tend to be shorter and to contain less semantic information~\cite{herka2023identifier, scanniello2017fixing}.  Therefore, even an expert human reader has less reason to read one word more than another as the program behavior must be understood from the interaction of  many different variables.%  In Java, key aspects of program behavior are more likely to be understood from single words, such as calls to a particular subroutine.  These language features are by design and not claims of this present paper~\citep{joy2000java, martin2002c, ogala2020comparative, zhu2015analysis} -- what we observe is likely an effect of these features.

Another possible explanation is the task the participants performed -- some research suggests task type has more effect on eye movement than language~\cite{mansoor2024assessing}.  The Smith study concerns locating bugs, which could require different patterns of skimming of the eyes~\cite{starke2009searching}.  Both the Wallace and Rodeghero concern writing code descriptions, with Wallace allowing navigation through code projects and Rodeghero limited to one function at a time.

\subsection{RQ2: Ablation Study}

We observed that our model with Pearson correlation outperforms the model without Pearson correlation. We show the results in Table~\ref{tab:shape-metrics}. One possible explanation is that Pearson correlation guides model to learn the linear relationship in the human attention. However, the increased performance of our model is not only due to Pearson correlation in the loss function coming from the three other metrics.
Spearman and Kendall are both higher in all cases, implying better \textit{rank} of words. Specifically, our model with Pearson correlation is more likely to produce a rank similar to that of a person's visual attention than baselines in a task of ranking words in source code.  The finding is corroborated by cosine similarity, which helps adjust for differences in total eye dwell time.  Some people may read more slowly, so they could have higher attention to all words they read, even if they pay more attention to the same words as other people do.  A higher similarity of these three metrics implies that our model is superior in terms of comparing words relative to each other.  This relative comparison may be more useful in practice than the absolute attention because software engineering tools often need to make binary decisions about what to show or not show~\cite{armaly2018comparison}.

\section{scanpath prediction}

\begin{table*}[b]
\centering
\small
\caption{Experimental results across different token size and approaches. The $\downarrow$ means the lower the number, the better the results. The $\uparrow$ means the larger the number, the better the results. Ours means our approach with Pearson correlation loss. The results are the average scores of function holdout. The numbers in bold means statistical significant difference between~\texttt{ours} and Bansal~\emph{et al.}~\cite{bansal2023modeling} with Wilcoxon test. The number in italic means the positive improvement \texttt{ours} and Bansal~\emph{et al.}~\cite{bansal2023modeling}}
\label{tab:token_results}
\begin{tabular}{lllcccc}
\hline
Task & Number of Tokens & Model & NLD $\downarrow$ & NDCG $\uparrow$ & Precision@k $\uparrow$ & Recall@k $\uparrow$ \\
\hline

\multirow{8}{*}{Reading}

& \multirow{4}{*}{5}

&  GPT-5 Nano    & 0.96 & 0.07 & 0.04 & 0.04\\
& & Claude-Haiku-4.5 & 0.96 & 0.08 & 0.05 & 0.05 \\
& &Bansal et al.~\cite{bansal2023modeling} & 0.89 & 0.23 & 0.19 & 0.19 \\
& & Ours          & \textit{\textbf{0.87}} & \textit{0.25} & \textit{0.22} & \textit{0.22} \\
\cline{2-7}

& \multirow{4}{*}{6}

&  GPT-5 Nano    & 0.96 &0.07 &0.04 & 0.04 \\
& & Claude-Haiku-4.5 & 0.96 & 0.07 &0.04 & 0.04\\
& & Bansal et al.~\cite{bansal2023modeling} & 0.89 & 0.23 & 0.20 & 0.20 \\
& & Ours          & \textbf{\textit{0.87}} & \textbf{\textit{0.27}} & \textbf{\textit{0.23}} & \textbf{\textit{0.23}} \\
\hline

\multirow{8}{*}{Writing}

& \multirow{4}{*}{5}

& GPT-5 Nano    & 0.95 &0.07 &0.04 & 0.04\\
& & Claude-Haiku-4.5 & 0.91 & 0.20 &0.17 &0.18 \\
& & Bansal et al.~\cite{bansal2023modeling} & 0.87 & 0.26 & 0.20 & 0.20 \\
& & Ours          & \textit{0.86} & \textit{0.28} & \textit{0.21} & \textit{0.22} \\
\cline{2-7}

& \multirow{4}{*}{6}

&  GPT-5 Nano    &0.96 & 0.06&0.03 &0.03 \\
& & Claude-Haiku-4.5 & 0.90 & 0.21 & 0.19 &0.19  \\
& & Bansal et al.~\cite{bansal2023modeling} & 0.88 & 0.27 & 0.22 & 0.22 \\
& & Ours          & \textbf{\textit{0.85}} & \textit{0.29} & 0.22 & 0.22 \\
\hline

\end{tabular}
\end{table*}

This section demonstrates the effectiveness of our models with scanpath prediction, which is a task of predicting a sequence of fixations 
\subsection{Research Questions}
Our research objective is to evaluate whether the model finetuned with the human attention improves the model's capability to solve the problems that require the understanding of human thought process. We use scanpath prediction as an example and ask the following Research Questions (RQs):
\vspace{1mm}
\begin{description}
    \item[~~RQ3] What is the performance of models finetuned with human attention in scanpath prediction compared with the baselines?
\end{description}

The rationale behind RQ3 is that the model finetuned with human attention may not improve the task that requires the understanding of human thought process. We compare our model (i.e., the mode trained with the Pearson correlation loss) with the baseline proposed by Bansal~\emph{et al.}~\cite{bansal2023modeling}.

\subsection{Dataset}
We use the dataset designed specifically for scanpath prediction proposed by Bansal~\emph{et al.}~\cite{bansal2023modeling}. They conducted an eye-tracking study with 27 programmers while doing the source code summarization. The study is conducted on the machine with Tobii Pro Fusion eye-tracker. Bansal~\emph{et al.}~\cite{bansal2023modeling} filter the noise by using velocity-based fixation and low-pass filter. This results in the scanpath for both reading source code and writing summary. We holdout one function each time and calculate the average score for each metrics. We do not use the data from Wallace study~\cite{wallace2025programmer} and Smith study~\cite{smith2025human} for scanpath prediction because the goal of both Smith study and Wallace study is to investigate how programmers understand source code to write a summary that describes the purpose of the function and localize bugs in the bigger context. These studies provide the participants the entire projects and investigate what context the programmers use to finish the task. Remember that the context size of the~\texttt{jam} model that we used is only 1,024, which is not possible to have the entire project as input. The investigation of the dataset that requires larger window size is beyond the scope of this paper. We do not consider Rodeghero study~\cite{rodeghero2014improving} because this dataset is an older dataset with smaller sample size. All things considered, we only use the dataset proposed by Bansal~\emph{et al.}~\cite{bansal2023modeling} because it aligns better with the current literature in scanpath prediction for software engineering tasks. 

\subsection{Metrics}
We used four different metrics based on the current literature, i.e., Normalized Levenshtein Distance (NLD),  Normalized Discounted Cumulative Gain (NDCG), Recall@k, and Precision@k. We provide our rationale for each as follows:
\begin{itemize}
    \item \textbf{Normalized Levenshtein Distance (NLD)} measures how many edits, adds, and substitutions are required to transform one string to another string. NLD normalized the value of Levenshtein distance to a range between zero and one by dividing it by the length of longest sequence, where zero means two sentences are identical and one means two sentence require the maximum edits, addition, or substitutions to transforms one sentence to another.
    \item \textbf{Normalized Disounted Cumulative Gain (NDCG)} measures whether  the model gives the highest rank to the most relevant tokens. The range of NDCG is between zero and one, where zero means the model does not demonstrate the ranking capability and one means the perfect ranking. 

    \item \textbf{Recall@k} measures how many  tokens are correct in the k predicted tokens. The range of recall@k is between zero and one, where zero means the model does not generate any correct tokens in k predicted tokens and one means all of the generated tokens are correct.

    \item \textbf{Precision@k} measures how many top k predicted tokens are correct. The range of precision@k is between zero and one, where zero means the model does not generate any correct tokens in k predicted tokens and one means top k generated tokens are correct.
\end{itemize}

Together, these metrics assess whether two strings are similar (NLD), the ranking capability of the model (NDCG), and whether the model generate the correct tokens (recall@k and precision@k).

\subsection{Baseline}
We used the baseline in software engineering proposed by Bansal~\emph{et al.}~\cite{bansal2023modeling} for comparison. We used this baseline because this baseline is most similar to our model. Bansal~\emph{et al.}~\cite{bansal2023modeling} modeled scanpath prediction as a finetuning task. They used the~\texttt{jam} model as their experiments. We used the same configurations and prompt as in the original paper for the baseline. We used the same prompt as in the baseline for our models. We also include two different close-sourced commercial models (i.e., Claude and GPT-5) as our baseline. We used the \texttt{claude-haiku-4-5-20251001} as our Claude model. We used \texttt{gpt-5-nano-2025-08-07
} as our GPT-5 model, which is the newest GPT-5 Nano model.

\subsection{Experiment Results}

We observed that our model significantly outperforms the baseline proposed by Bansal~\emph{et al.}~\cite{bansal2023modeling} in NLD on the reading task. We show the results in Table~\ref{tab:token_results}. We also observed the positive improvement in NDCG, precision@k, and recall@k. For example, compared with the baseline proposed by Bansal~\emph{et al.}~\cite{bansal2023modeling}, we observed 8\% and 17\% improvement in NDCG for five and six tokens prediction tasks respectively.  This shows that our model is able to generate the scanpath similar to the human programmers and rank the most relevant tokens in the highest positions. One possible explanation is that our model has already been finetuned with human visual attention. This finetuning provides the model with guidance on  where to focus in the software engineering tasks. We also observed that the performance of those commercial models (i.e, GPT-5 and Claude) is even worse than our baseline. One of the possible explanations is that those models are trained with large-scale data from the Internet. However, those commercial models could  look at different tokens to do software engineering tasks compared with human programmers~\cite{paltenghi2021thinking} and demonstrate an alien thought process~\cite{su2026code}. Overall, we observed that our model is more capable of the task required the understanding of human thought process.

We observed a positive improvement on the writing task compared with Bansal baseline. We show the result in Table~\ref{tab:token_results}. Specifically, we observed a 2\% improvement in NLD and a 7\% improvement in NDCG on the six-token prediction task although the differences between \texttt{ours} and the baseline proposed by Bansal~\emph{et al.} were not statistically significant. A possible explanation is that human programmers could have the different visual attention in the writing task  compared to the reading task. In fact, Karas~\emph{et al.}~\cite{karas2024tale} found the farther reading distance when programmers are doing writing tasks. Overall, our results still show the positive improvement over commercial models and the Bansal baseline on the writing task although our model is only trained on visual attention during source code comprehension. 

We show one reading example in Table~\ref{tab:reading_example}. We observed that all of the commercial models (GPT-5 and Claude) first focus on the words in the function signature. For example, we observed that the first token of those commercial models is genSql. This aligns with the current literature that function signature provide more information for source code summarization models~\cite{ding2024code}. However, we observe that the first token that the human programmer look is SqlcPrettyPrinter, which is in the body of the function. Our model is able to predict SqlcPrettyPrinter as the first token and public as the last token, which aligns better with the human programmers. We also observe that Bansal baseline is not able to genearte the correct tokens. Overall, we found our model finetuned with human attention generates a scanpath closer to the scanpath of human programmers although we also caution that our model is more human-like compared with baselines.

% need to use 311_29852582
\begin{table}[t!]
\centering
\small
\vspace{-2mm}
\caption{Code summarization example for Java }
\label{tab:reading_example}
\begin{tabular}{ll}
\textbf Method ID 311\_29852582 \\\hline
\end{tabular}

\begin{tabular}{c}
\begin{lstlisting}[language=Java, basicstyle=\ttfamily\small, breaklines=true, showstringspaces=false]
 public void genSql() throws PositionedError {
    try {
      SqlcPrettyPrinter         spp;

      spp = new SqlcPrettyPrinter(ref.getFile());
      spp.printCUnit(elems);
      spp.close();
    } catch (IOException ioe) {
      ioe.printStackTrace();
      System.err.println("cannot write: " + ref.getFile());
    }
  }
 
\end{lstlisting} \\

\end{tabular}
\begin{tabular}{l|p{5cm}}
\hline
     Bansal~\emph{et al.}~\cite{bansal2023modeling} & Jexpression[] TokenReference TokenReference ref getTokenReference JExpression.1 TokenReference\\\hline
     GPT-5 Nano & genSql	throws	PositionedError	try	SqlcPrettyPrinter	spp \\\hline
     Claude-Haiku-4.5 &genSql	void	SqlcPrettyPrinter	printCUnit	IOException	catch \\\hline
     Ours & SqlcPrettyPrinter public void SqlcPrettyPrinter new public \\\hline
     reference& SqlcPrettyPrinter spp.printCUnit void SqlcPrettyPrinter try public
\end{tabular}
\end{table}
\section{threat to validity}
We follow Wohlin~\emph{et al.}~\cite{wohlin2012experimentation} to organize threat to validity into threat to internal validity, external validity, and construct validity.  The threat to internal validity is that the baseline that we compare may not be a fair comparison. We mitigate this threat by including the baseline from both computer vision and software engineering. We mitigate this threat on scanpath prediction by including both the baseline from the current literature and the strong commercial models as our baseline. The threat to external validity is that our model might not be able to generalize to other software engineering tasks and programming languages. We mitigate this threat by using diverse dataset. We use the dataset from two different tasks and two different programming languages although it is possible that our model does not generalize to tasks other than source code summarization and bug localization. The threat to construct validity is that the metrics that we use may not be appropriate to measure our tasks. We mitigate this threat by using the metrics from the current literature. 

\section{limitation}
First, our supervision signal, \textit{ptgt}, aggregates visual attention across time and collapses reading order and eye movement regressions into a single scalar per token~\cite{bansal2024modeling, rodeghero2014improving}.  As a result, our model captures where attention is allocated but not how attention may change over time.  Second, we caution that we do not intend to take a side in the ``attention equals explanation'' debate~\cite{bibal2022attention, jain2019attention, wiegreffe2019attention} even though our model design may be characterized as a ``surgical intervention''~\cite{grimsley2020attention}.  We are studying human and machine attention itself in this paper, not its downstream effect and we do not claim that our model implements the same mechanisms that give rise to human attention (our model is not a surrogate human). We also caution that our results may or may not generalize over other datasets, languages, or tasks.  Our model is not a complete account of all aspects of attention, but is nonetheless a step forward in computational modeling of human programmer visual attention. Finally, we caution that our results for scanpath may not be able to generalize to a dataset that requires large context due to our model size.

\begin{figure}[t]
    \centering

    % Main figure
    \includegraphics[width=1.0\columnwidth]{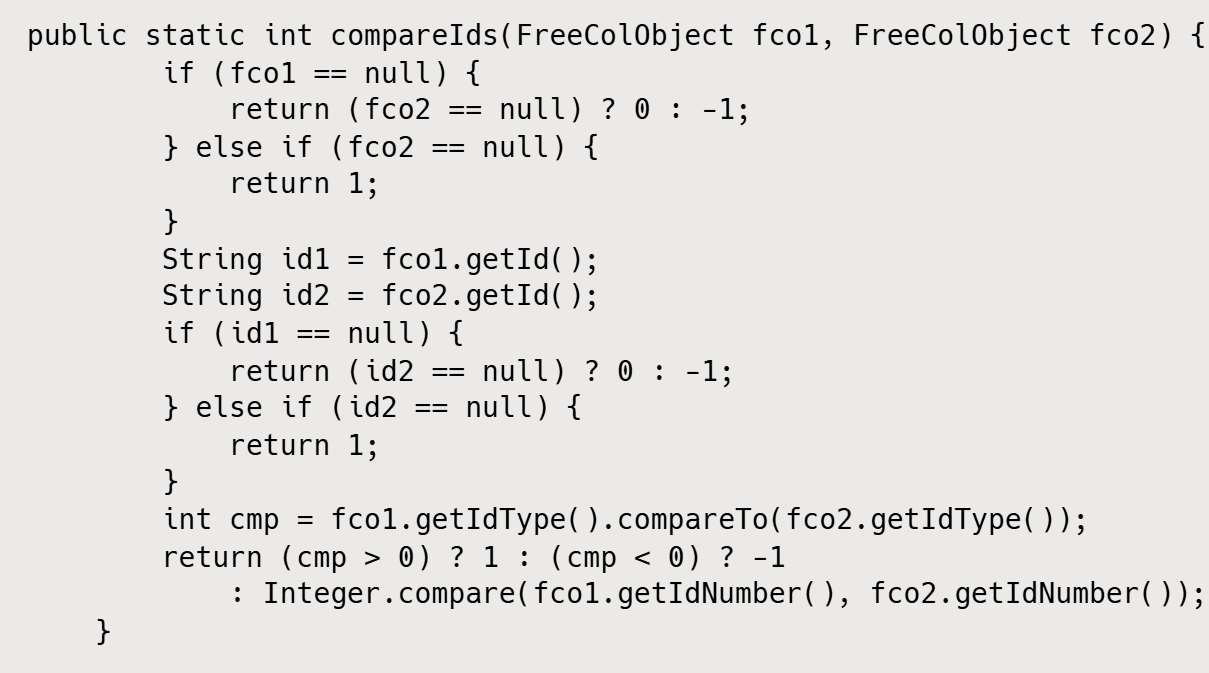}

    \vspace{2mm}

    \subfloat[Aggregate Reference Human Attention\label{fig:human_reference}]{
        \includegraphics[width=1.0\columnwidth]{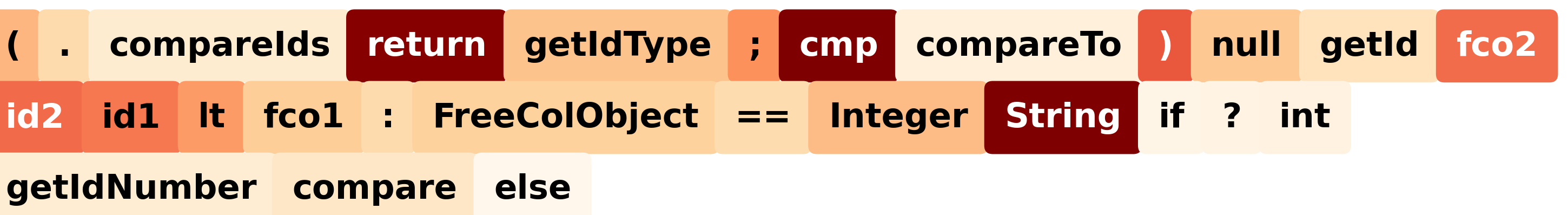}
    }

    \vspace{2mm}

    \subfloat[Predicted Attention from \texttt{ours-nocorr}\label{fig:ours_nocorr}]{
        \includegraphics[width=1.0\columnwidth]{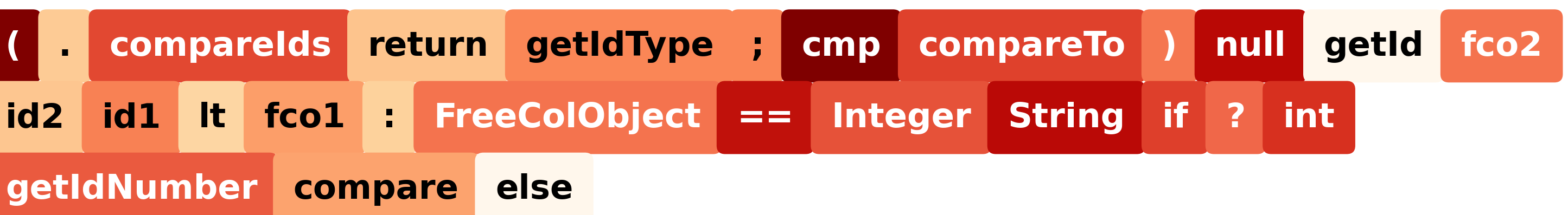}
    }

    \vspace{2mm}

    \subfloat[Predicted Attention from \texttt{ours}\label{fig:ours}]{
        \includegraphics[width=1.0\columnwidth]{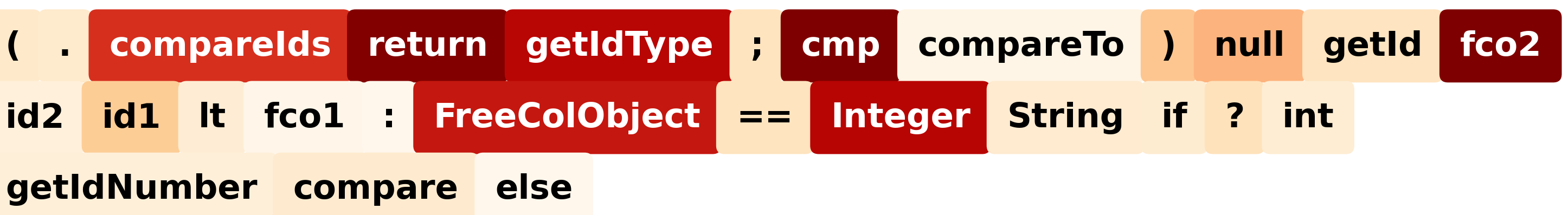}
    }

    \caption{Example Java method \texttt{compareIds()} from the Wallace study followed by heatmaps of aggregate reference human attention and predicted attention from the main and nocorr versions of our model. Darker shades of red indicate higher levels of actual or predicted \textit{ptgt}.}
    \label{fig:attention_comparison}
\end{figure}

\section{Discussion}

A central finding of this work is that a Transformer language model can predict aggregate human visual attention over source code with a level comparable to that of individual human participants relative to their group when we finetuned with eye-tracking supervision.  This finding reflects the well-known phenomenon that averaging across observers reveals stable attentional structure even if that structure is partially obscured by individual variability.  In our results, this effect is measurable when comparing machine-human similarity and the human reference baseline.  Individual programmers exhibit idiosyncratic reading strategies, but they form a consensus on overall need to attend to certain words. This finding is consistent with studies comparing, for example, word-level attention of blind and sighted programmers~\cite{armaly2018comparison}.

We demonstrate a model that is optimized against an aggregate attention signal (percent total gaze time).  In effect the model learns a population-level attentional prior over tokens in software source code. In this way, the model should be viewed as an estimator of aggregate attentional structure across programmers rather than as a surrogate human.  Consider the example in Figure~\ref{fig:attention_comparison}.  Subfigure (a) of the example shows the aggregate \textit{ptgt} we calculate, where darker shades of red indicate more attention.  The list of words includes all which were read by any person in the source code for the Java method \texttt{compareIds()} above.  Consider the token \texttt{compareTo}. This token is the name of a method in a method invocation.  Method invocations are critical to program behavior.  There is a long-observed tendency of artificial neural networks to attend to method invocations due to their position as connection points among source code components and co-occurrence with other words~\cite{chirkova2021empirical, gao2023code, mohammadkhani2023explaining, sharma2022exploratory}.  Thus, it is not surprising that \texttt{ours-nocorr} predicted relatively high attention to this word, visible as medium-dark red highlighting in Figure~\ref{fig:attention_comparison}b.

Nevertheless, the aggregate observed human attention to \texttt{compareTo} is low (Figure~\ref{fig:attention_comparison}a).  The model labeled \texttt{ours} correctly predicted low attention to that word (Figure~\ref{fig:attention_comparison}c).  The possible explanation is that this model received the \textit{ptgt} visual attention as a signal during training.  The aggregate attention paid to method names in an invocation is actually fairly low.  In fact, \cite{abid2019developer} found that programmers typically read the \textit{inputs} to the invocation much more than the invoked method's name, a finding also consistent with the high attention prediction for the word \texttt{fco2}, which is an input to the \texttt{compareTo} method.  Therefore, even though neural models tend to have high attention to method invocation names, this high attention is inconsistent with human visual attention, and our model's prediction was in closer alignment with this human reference.

From a technical perspective, one of the more consequential design decisions in our model is to extract attention from the final Transformer layer rather than training a new output layer specifically to generate attention predictions.  This decision allowed us to keep the model's architecture intact and instead only use a custom loss function during finetuning.  The model $\mathcal{M}^{\text{ft}}$ we create has the same autoregressive behavior as $\mathcal{M}^{\text{pre}}$.  Therefore our model retains the ability to respond to prompts.  An intriguing possibility, hinted to by related work~\cite{bansal2023towards}, is that the model $\mathcal{M}^{\text{ft}}$ may exhibit more human-like responses to prompts than $\mathcal{M}^{\text{pre}}$ because its attention mechanism is more human-like. We showed that our model has positive improvement over scanpath prediction tasks, which is a task that requires the understanding of human thought process. %However, investigation of this possibility is beyond the scope of this paper and a subject of our immediate future work.

\section{Conclusion}
We move the stat-of-the-art threefold. First, we develop a novel loss function to train a model with the data from the eye-tracking studies. We used two programming languages (Java and C) and two programming tasks (bug localization and source code summarization) as examples. We showed that our model outperforms the close baselines in both computer vision and software engineering. 

Second, we conducted an ablation to study to investigate the impact of the loss function with Pearson correlation. We found that Pearson correlation guides models to learn the linear relationship in the data from the eye-tracking studies. We also show that Pearson correlation is not the only contribution to the performance based on our experiments results with other metrics (i.e., Kendall’s $\tau$, Spearman rank correlation, and Cosine similarity).

Finally, we demonstrated that our model is  more capable of the tasks that require the understanding of human thought process. We used scanpath prediction as examples. We show the improvement in both reading and writing tasks compared with the close baseline and commercial models (i.e., GPT-5 and Claude). We also caution that our model may not be able to generalize to a task with large context windows size.

We release the source code in the online appendix at \url{https://github.com/apcl-research/human-attention-prediction}.

\section*{ackowledgment}
This work is supported in part by the NSF grants CCF-2100035, CCF-2211428, and CCF-2211429. Any opinions, findings, and conclusions expressed herein are the authors’ and do not necessarily reflect those of the sponsors. We also sincerely thank participants of our qualitative study.

\bibliographystyle{IEEEtran}
\bibliography{ref}

@article{wallace2025programmer,
author = {Wallace, Robert and Bansal, Aakash and Karas, Zachary and Tang, Ningzhi and Huang, Yu and Jia-Jun Li, Toby and McMillan, Collin},
title = {Programmer Visual Attention During Context-Aware Code Summarization},
year = {2025},
issue_date = {May 2025},
publisher = {IEEE Press},
volume = {51},
number = {5},
issn = {0098-5589},
url = {https://doi.org/10.1109/TSE.2025.3554990},
doi = {10.1109/TSE.2025.3554990},
journal = {IEEE Trans. Softw. Eng.},
month = may,
pages = {1524–1537},
numpages = {14}
}

@inproceedings{rodeghero2014improving,
author = {Rodeghero, Paige and McMillan, Collin and McBurney, Paul W. and Bosch, Nigel and D'Mello, Sidney},
title = {Improving automated source code summarization via an eye-tracking study of programmers},
year = {2014},
isbn = {9781450327565},
publisher = {Association for Computing Machinery},
address = {New York, NY, USA},
url = {https://doi.org/10.1145/2568225.2568247},
doi = {10.1145/2568225.2568247},
booktitle = {Proceedings of the 36th International Conference on Software Engineering},
pages = {390–401},
numpages = {12},
location = {Hyderabad, India},
series = {ICSE 2014}
}

@inproceedings{saha2025modeling,
  title     = {Modeling the Human Visual System: Comparative Insights from Response-Optimized and Task-Optimized Vision Models, Language Models, and Different Readout Mechanisms},
  author    = {Saha, Shreya and Chadha, Ishaan and Khosla, Meenakshi},
  booktitle = {Proceedings of the 8th Annual Conference on Cognitive Computational Neuroscience (CCN 2025)},
  year      = {2025},
  publisher = {Cognitive Computational Neuroscience},
  url       = {https://openreview.net/forum?id=QA0P53hQRT}
}

@article{smith2025human,
  title={Human Attention During Localization of Memory Bugs in C Programs},
  author={Smith, Emory and Wallace, Robert and Robison, Matthew and Huang, Yu and McMillan, Collin},
  journal={arXiv preprint arXiv:2506.00693},
  year={2025}
}

@article{carrasco2011visual,
  title={Visual attention: The past 25 years},
  author={Carrasco, Marisa},
  journal={Vision research},
  volume={51},
  number={13},
  pages={1484--1525},
  year={2011},
  publisher={Elsevier}
}

@article{cheng2012eye,
  title={Eye-tracking based adaptive user interface: implicit human-computer interaction for preference indication},
  author={Cheng, Shiwei and Liu, Ying},
  journal={Journal on Multimodal User Interfaces},
  volume={5},
  number={1},
  pages={77--84},
  year={2012},
  publisher={Springer}
}

@article{von2002program,
  title={Program comprehension during software maintenance and evolution},
  author={Von Mayrhauser, Anneliese and Vans, A Marie},
  journal={Computer},
  volume={28},
  number={8},
  pages={44--55},
  year={2002},
  publisher={IEEE}
}

@inproceedings{bouraffa2023developers,
  title={Developers' Visuo-spatial Mental Model and Program Comprehension},
  author={Bouraffa, Abir and Fuhrmann, Gian-Luca and Maalej, Walid},
  booktitle={2023 IEEE/ACM 45th International Conference on Software Engineering (ICSE)},
  pages={1920--1932},
  year={2023},
  organization={IEEE}
}

@article{calvo2010affect,
  title={Affect detection: An interdisciplinary review of models, methods, and their applications},
  author={Calvo, Rafael A and D'Mello, Sidney},
  journal={IEEE Transactions on affective computing},
  volume={1},
  number={1},
  pages={18--37},
  year={2010},
  publisher={IEEE}
}

@inproceedings{su2023language,
  title={A language model of java methods with train/test deduplication},
  author={Su, Chia-Yi and Bansal, Aakash and Jain, Vijayanta and Ghanavati, Sepideh and McMillan, Collin},
  booktitle={Proceedings of the 31st ACM Joint European Software Engineering Conference and Symposium on the Foundations of Software Engineering},
  pages={2152--2156},
  year={2023}
}

@inproceedings{xu2022systematic,
  title={A systematic evaluation of large language models of code},
  author={Xu, Frank F and Alon, Uri and Neubig, Graham and Hellendoorn, Vincent Josua},
  booktitle={Proceedings of the 6th ACM SIGPLAN international symposium on machine programming},
  pages={1--10},
  year={2022}
}

@article{naveed2025comprehensive,
  title={A comprehensive overview of large language models},
  author={Naveed, Humza and Khan, Asad Ullah and Qiu, Shi and Saqib, Muhammad and Anwar, Saeed and Usman, Muhammad and Akhtar, Naveed and Barnes, Nick and Mian, Ajmal},
  journal={ACM Transactions on Intelligent Systems and Technology},
  volume={16},
  number={5},
  pages={1--72},
  year={2025},
  publisher={ACM New York, NY}
}

@inproceedings{wang2024probing,
  title={Probing large language models from a human behavioral perspective},
  author={Wang, Xintong and Li, Xiaoyu and Li, Xingshan and Biemann, Chris},
  booktitle={Proceedings of the Workshop: Bridging Neurons and Symbols for Natural Language Processing and Knowledge Graphs Reasoning (NeusymBridge)@ LREC-COLING-2024},
  pages={1--7},
  year={2024}
}

@article{letovsky1987cognitive,
  title={Cognitive processes in program comprehension},
  author={Letovsky, Stanley},
  journal={Journal of Systems and software},
  volume={7},
  number={4},
  pages={325--339},
  year={1987},
  publisher={Elsevier}
}

@article{sharafi2015systematic,
  title={A systematic literature review on the usage of eye-tracking in software engineering},
  author={Sharafi, Zohreh and Soh, Z{\'e}phyrin and Gu{\'e}h{\'e}neuc, Yann-Ga{\"e}l},
  journal={Information and Software Technology},
  volume={67},
  pages={79--107},
  year={2015},
  publisher={Elsevier}
}

@inproceedings{eberle2022transformer,
  title={Do transformer models show similar attention patterns to task-specific human gaze?},
  author={Eberle, Oliver and Brandl, Stephanie and Pilot, Jonas and S{\o}gaard, Anders},
  booktitle={Proceedings of the 60th Annual Meeting of the Association for Computational Linguistics (Volume 1: Long Papers)},
  pages={4295--4309},
  year={2022}
}

@article{davoudi2015critical,
  title={Critical review of the models of reading comprehension with a focus on situation models},
  author={Davoudi, Mohammad and Moghadam, Hamid Reza Hashemi},
  journal={International Journal of Linguistics},
  volume={7},
  number={5},
  pages={172--187},
  year={2015}
}

@article{rayner2010models,
  title={Models of the reading process},
  author={Rayner, Keith and Reichle, Erik D},
  journal={Wiley Interdisciplinary Reviews: Cognitive Science},
  volume={1},
  number={6},
  pages={787--799},
  year={2010},
  publisher={Wiley Online Library}
}

@inproceedings{mazzeo2021deep,
  title={Deep learning based eye gaze estimation and prediction},
  author={Mazzeo, Pier Luigi and D'Amico, Dilan and Spagnolo, Paolo and Distante, Cosimo},
  booktitle={2021 6th International Conference on Smart and Sustainable Technologies (SpliTech)},
  pages={1--6},
  year={2021},
  organization={IEEE}
}

@inproceedings{zhao2016gaze,
  title={Gaze prediction for recommender systems},
  author={Zhao, Qian and Chang, Shuo and Harper, F Maxwell and Konstan, Joseph A},
  booktitle={Proceedings of the 10th ACM Conference on Recommender Systems},
  pages={131--138},
  year={2016}
}

@article{hu2020dgaze,
  title={Dgaze: Cnn-based gaze prediction in dynamic scenes},
  author={Hu, Zhiming and Li, Sheng and Zhang, Congyi and Yi, Kangrui and Wang, Guoping and Manocha, Dinesh},
  journal={IEEE transactions on visualization and computer graphics},
  volume={26},
  number={5},
  pages={1902--1911},
  year={2020},
  publisher={IEEE}
}

@article{littman1987mental,
  title={Mental models and software maintenance},
  author={Littman, David C and Pinto, Jeannine and Letovsky, Stanley and Soloway, Elliot},
  journal={Journal of Systems and Software},
  volume={7},
  number={4},
  pages={341--355},
  year={1987},
  publisher={Elsevier}
}

@inproceedings{bansal2023modeling,
  title={Modeling programmer attention as scanpath prediction},
  author={Bansal, Aakash and Su, Chia-Yi and Karas, Zachary and Zhang, Yifan and Huang, Yu and Li, Toby Jia-Jun and McMillan, Collin},
  booktitle={2023 38th IEEE/ACM International Conference on Automated Software Engineering (ASE)},
  pages={1732--1736},
  year={2023},
  organization={IEEE}
}

@article{karas2024tale,
  title={A tale of two comprehensions? analyzing student programmer attention during code summarization},
  author={Karas, Zachary and Bansal, Aakash and Zhang, Yifan and Li, Toby and McMillan, Collin and Huang, Yu},
  journal={ACM Transactions on Software Engineering and Methodology},
  volume={33},
  number={7},
  pages={1--37},
  year={2024},
  publisher={ACM New York, NY}
}

@inproceedings{abid2019using,
author = {Abid, Nahla J. and Maletic, Jonathan I. and Sharif, Bonita},
title = {Using developer eye movements to externalize the mental model used in code summarization tasks},
year = {2019},
isbn = {9781450367097},
publisher = {Association for Computing Machinery},
address = {New York, NY, USA},
url = {https://doi.org/10.1145/3314111.3319834},
doi = {10.1145/3314111.3319834},
booktitle = {Proceedings of the 11th ACM Symposium on Eye Tracking Research \& Applications},
articleno = {13},
numpages = {9},
location = {Denver, Colorado},
series = {ETRA '19}
}

@ARTICLE{sharafi2022eyes,
  author={Sharafi, Zohreh and Bertram, Ian and Flanagan, Michael and Weimer, Westley},
  journal={IEEE Transactions on Software Engineering}, 
  title={Eyes on Code: A Study on Developers’ Code Navigation Strategies}, 
  year={2022},
  volume={48},
  number={5},
  pages={1692-1704},
  doi={10.1109/TSE.2020.3032064}}

@article{kevic2017eye,
  title={Eye gaze and interaction contexts for change tasks--observations and potential},
  author={Kevic, Katja and Walters, Braden M and Shaffer, Timothy R and Sharif, Bonita and Shepherd, David C and Fritz, Thomas},
  journal={Journal of Systems and Software},
  volume={128},
  pages={252--266},
  year={2017},
  publisher={Elsevier}
}

@article{grandel2025comcat,
  title={ComCat: Expertise-Guided Context Generation to Enhance Code Comprehension},
  author={Grandel, Skyler and Andersen, Scott Thomas and Huang, Yu and Leach, Kevin},
  journal={ACM Transactions on Software Engineering and Methodology},
year={2025},
  publisher={ACM New York, NY}
}

@inproceedings{wiegreffe2019attention,
  title={Attention is not not Explanation},
  author={Wiegreffe, Sarah and Pinter, Yuval},
  booktitle={Proceedings of the 2019 Conference on Empirical Methods in Natural Language Processing and the 9th International Joint Conference on Natural Language Processing (EMNLP-IJCNLP)},
  pages={11--20},
  year={2019}
}

@inproceedings{grimsley2020attention,
  title={Why attention is not explanation: Surgical intervention and causal reasoning about neural models},
  author={Grimsley, Christopher and Mayfield, Elijah and Bursten, Julia RS},
  booktitle={Proceedings of the Twelfth Language Resources and Evaluation Conference},
  pages={1780--1790},
  year={2020}
}

@inproceedings{bibal2022attention,
  title={Is attention explanation? an introduction to the debate},
  author={Bibal, Adrien and Cardon, R{\'e}mi and Alfter, David and Wilkens, Rodrigo and Wang, Xiaoou and Fran{\c{c}}ois, Thomas and Watrin, Patrick},
  booktitle={Proceedings of the 60th Annual Meeting of the Association for Computational Linguistics (Volume 1: Long Papers)},
  pages={3889--3900},
  year={2022}
}

@article{saha2024modeling,
  title={Modeling the human visual system: Comparative insights from response-optimized and task-optimized vision models, language models, and different readout mechanisms},
  author={Saha, Shreya and Chadha, Ishaan and Khosla, Meenakshi},
  journal={arXiv preprint arXiv:2410.14031},
  year={2024}
}

@inproceedings{wang2025cognitive,
  title={Cognitive Insights into Document Comprehension: The Role of Reading Order and Visual Attention in Human and Large Language Models},
  author={Wang, QingXuan and Wang, Hao and Zhang, Huiran and Chu, Chenhui and Wang, Rui and Zhu, Pinpin},
  booktitle={Proceedings of the Annual Meeting of the Cognitive Science Society},
  volume={47},
  year={2025}
}

@inproceedings{jain2019attention,
  title={Attention is not Explanation},
  author={Jain, Sarthak and Wallace, Byron C},
  booktitle={Proceedings of the 2019 Conference of the North American Chapter of the Association for Computational Linguistics: Human Language Technologies, Volume 1 (Long and Short Papers)},
  pages={3543--3556},
  year={2019}
}

@inproceedings{abid2019developer,
  title={Developer reading behavior while summarizing java methods: Size and context matters},
  author={Abid, Nahla J and Sharif, Bonita and Dragan, Natalia and Alrasheed, Hend and Maletic, Jonathan I},
  booktitle={2019 IEEE/ACM 41st International Conference on Software Engineering (ICSE)},
  pages={384--395},
  year={2019},
  organization={IEEE}
}

@inproceedings{mohammadkhani2023explaining,
  title={Explaining transformer-based code models: What do they learn? when they do not work?},
  author={Mohammadkhani, Ahmad Haji and Tantithamthavorn, Chakkrit and Hemmatif, Hadi},
  booktitle={2023 IEEE 23rd International Working Conference on Source Code Analysis and Manipulation (SCAM)},
  pages={96--106},
  year={2023},
  organization={IEEE}
}

@article{gao2023code,
  title={Code structure--guided transformer for source code summarization},
  author={Gao, Shuzheng and Gao, Cuiyun and He, Yulan and Zeng, Jichuan and Nie, Lunyiu and Xia, Xin and Lyu, Michael},
  journal={ACM Transactions on Software Engineering and Methodology},
  volume={32},
  number={1},
  pages={1--32},
  year={2023},
  publisher={ACM New York, NY}
}

@inproceedings{chirkova2021empirical,
  title={Empirical study of transformers for source code},
  author={Chirkova, Nadezhda and Troshin, Sergey},
  booktitle={Proceedings of the 29th ACM joint meeting on European software engineering conference and symposium on the foundations of software engineering},
  pages={703--715},
  year={2021}
}

@inproceedings{sharma2022exploratory,
  title={An exploratory study on code attention in BERT},
  author={Sharma, Rishab and Chen, Fuxiang and Fard, Fatemeh and Lo, David},
  booktitle={Proceedings of the 30th ieee/acm international conference on program comprehension},
  pages={437--448},
  year={2022}
}

@inproceedings{sun2023program,
  title={Program Code Navigation Model for Individuals based on LSTM with Co-clustering},
  author={Sun, Ming and Nakayama, Minoru},
  booktitle={Proceedings of the 2023 Symposium on Eye Tracking Research and Applications},
  pages={1--6},
  year={2023}
}

@article{hasson2010reliability,
  title={Reliability of cortical activity during natural stimulation},
  author={Hasson, Uri and Malach, Rafael and Heeger, David J},
  journal={Trends in cognitive sciences},
  volume={14},
  number={1},
  pages={40--48},
  year={2010},
  publisher={Elsevier}
}

@article{mansoor2024assessing,
  title={Assessing the effect of programming language and task type on eye movements of computer science students},
  author={Mansoor, Niloofar and Peterson, Cole S and Dodd, Michael D and Sharif, Bonita},
  journal={ACM Transactions on Computing Education},
  volume={24},
  number={1},
  pages={1--38},
  year={2024},
  publisher={ACM New York, NY}
}

@misc{nastase2019measuring,
  title={Measuring shared responses across subjects using intersubject correlation},
  author={Nastase, Samuel A and Gazzola, Valeria and Hasson, Uri and Keysers, Christian},
  journal={Social cognitive and affective neuroscience},
  volume={14},
  number={6},
  pages={667--685},
  year={2019},
  publisher={Oxford University Press}
}

@article{hasson2004intersubject,
  title={Intersubject synchronization of cortical activity during natural vision},
  author={Hasson, Uri and Nir, Yuval and Levy, Ifat and Fuhrmann, Galit and Malach, Rafael},
  journal={science},
  volume={303},
  number={5664},
  pages={1634--1640},
  year={2004},
  publisher={American Association for the Advancement of Science}
}

@inproceedings{starke2009searching,
  title={Searching and skimming: An exploratory study},
  author={Starke, Jamie and Luce, Chris and Sillito, Jonathan},
  booktitle={2009 IEEE international conference on software maintenance},
  pages={157--166},
  year={2009},
  organization={IEEE}
}

@article{herka2023identifier,
  title={Identifier Names in Computer Programs: Literature Review.},
  author={Herka, Iwo},
  journal={Advances in Cognitive Psychology},
  volume={19},
  number={3},
  year={2023}
}

@article{scanniello2017fixing,
  title={Fixing faults in c and java source code: Abbreviated vs. full-word identifier names},
  author={Scanniello, Giuseppe and Risi, Michele and Tramontana, Porfirio and Romano, Simone},
  journal={ACM Transactions on Software Engineering and Methodology (TOSEM)},
  volume={26},
  number={2},
  pages={1--43},
  year={2017},
  publisher={ACM New York, NY, USA}
}

@inproceedings{armaly2018comparison,
  title={A comparison of program comprehension strategies by blind and sighted programmers},
  author={Armaly, Ameer and Rodeghero, Paige and McMillan, Collin},
  booktitle={Proceedings of the 40th International Conference on Software Engineering},
  pages={788--788},
  year={2018}
}

@article{xu2014predicting,
  title={Predicting human gaze beyond pixels},
  author={Xu, Juan and Jiang, Ming and Wang, Shuo and Kankanhalli, Mohan S and Zhao, Qi},
  journal={Journal of vision},
  volume={14},
  number={1},
  pages={28--28},
  year={2014},
  publisher={The Association for Research in Vision and Ophthalmology}
}

@article{tafasca2024toward,
  title={Toward semantic gaze target detection},
  author={Tafasca, Samy and Gupta, Anshul and Bros, Victor and Odobez, Jean-Marc},
  journal={Advances in neural information processing systems},
  volume={37},
  pages={121422--121448},
  year={2024}
}

@article{su2024distilled,
  title={Distilled GPT for source code summarization},
  author={Su, Chia-Yi and McMillan, Collin},
  journal={Automated Software Engineering},
  volume={31},
  number={1},
  pages={22},
  year={2024},
  publisher={Springer}
}

@inproceedings{McLoughlin2025VisualAttentionFCG,
  author    = {Samantha McLoughlin and Zachary Karas and Robert Wallace and Aakash Bansal and Collin McMillan and Yu Huang},
  title     = {Programmers' Visual Attention on Function Call Graphs During Code Summarization},
  booktitle = {Proceedings of the 40th IEEE/ACM International Conference on Automated Software Engineering (ASE '25)},
  year      = {2025},
  month     = nov,
  address   = {Nov. 16--20},
  publisher = {IEEE/ACM}
}

@inproceedings{li2024machines,
  title={Do machines and humans focus on similar code? exploring explainability of large language models in code summarization},
  author={Li, Jiliang and Zhang, Yifan and Karas, Zachary and McMillan, Collin and Leach, Kevin and Huang, Yu},
  booktitle={Proceedings of the 32nd IEEE/ACM International Conference on Program Comprehension},
  pages={47--51},
  year={2024}
}

@inproceedings{yoshioka2025eye2vec,
  title={eye2vec: Learning Distributed Representations of Eye Movement for Program Comprehension Analysis},
  author={Yoshioka, Haruhiko and Shimari, Kazumasa and Uwano, Hidetake and Matsumoto, Kenichi},
  booktitle={Proceedings of the 2025 Symposium on Eye Tracking Research and Applications},
  pages={1--3},
  year={2025}
}

@inproceedings{abbad2022estimating,
  title={Estimating developers' cognitive load at a fine-grained level using eye-tracking measures},
  author={Abbad-Andaloussi, Amine and Sorg, Thierry and Weber, Barbara},
  booktitle={Proceedings of the 30th IEEE/ACM international conference on program comprehension},
  pages={111--121},
  year={2022}
}

@inproceedings{de2020toward,
  title={Toward a definition of cognitive-driven development},
  author={de Souza, Alberto Luiz Oliveira Tavares and Pinto, Victor Hugo Santiago Costa},
  booktitle={2020 IEEE International Conference on Software Maintenance and Evolution (ICSME)},
  pages={776--778},
  year={2020},
  organization={IEEE}
}

@inproceedings{wong20223d,
  title={3D segmentation with fully trainable Gabor kernels and Pearson’s correlation coefficient},
  author={Wong, Ken CL and Moradi, Mehdi},
  booktitle={International Workshop on Machine Learning in Medical Imaging},
  pages={53--61},
  year={2022},
  organization={Springer}
}

@inproceedings{bansal2024modeling,
author = {Bansal, Aakash and Su, Chia-Yi and Karas, Zachary and Zhang, Yifan and Huang, Yu and Li, Toby Jia-Jun and McMillan, Collin},
title = {Modeling Programmer Attention as Scanpath Prediction},
year = {2024},
isbn = {9798350329964},
publisher = {IEEE Press},
url = {https://doi.org/10.1109/ASE56229.2023.00092},
doi = {10.1109/ASE56229.2023.00092},
booktitle = {Proceedings of the 38th IEEE/ACM International Conference on Automated Software Engineering},
pages = {1732–1736},
numpages = {5},
location = {Echternach, Luxembourg},
series = {ASE '23}
}

@inproceedings{kuang2023applying,
  title={Applying Machine Learning to Gaze Data in Software Development: a Mapping Study},
  author={Kuang, Peng and S{\"o}derberg, Emma and Niehorster, Diederick C and H{\"o}st, Martin},
  booktitle={Proceedings of the 2023 Symposium on Eye Tracking Research and Applications},
  pages={1--7},
  year={2023}
}

@article{paltenghi2024follow,
  title={Follow-up Attention: An Empirical Study of Developer and Neural Model Code Exploration},
  author={Paltenghi, Matteo and Pandita, Rahul and Henley, Austin Z and Ziegler, Albert},
  journal={IEEE Transactions on Software Engineering},
  year={2024},
  publisher={IEEE}
}

@article{sharif2017eye,
  title={Eye movements in software traceability link recovery},
  author={Sharif, Bonita and Meinken, John and Shaffer, Timothy and Kagdi, Huzefa},
  journal={Empirical Software Engineering},
  volume={22},
  number={3},
  pages={1063--1102},
  year={2017},
  publisher={Springer}
}

@article{zhang2025eyemulator,
  title={EyeMulator: Improving Code Language Models by Mimicking Human Visual Attention},
  author={Zhang, Yifan and Huang, Chen and Zhang, Yueke and Zhang, Jiahao and Li, Toby Jia-Jun and McMillan, Collin and Leach, Kevin and Huang, Yu},
  journal={arXiv preprint arXiv:2508.16771},
  year={2025}
}

@inproceedings{konopka2015combining,
  title={Combining eye tracking with navigation paths for identification of cross-language code dependencies},
  author={Konopka, Martin},
  booktitle={Proceedings of the 2015 10th Joint Meeting on Foundations of Software Engineering},
  pages={1057--1059},
  year={2015}
}

@article{obaidellah2018survey,
  title={A survey on the usage of eye-tracking in computer programming},
  author={Obaidellah, Unaizah and Al Haek, Mohammed and Cheng, Peter C-H},
  journal={ACM Computing Surveys (CSUR)},
  volume={51},
  number={1},
  pages={1--58},
  year={2018},
  publisher={ACM New York, NY, USA}
}

@article{sharafi2020practical,
  title={A practical guide on conducting eye tracking studies in software engineering},
  author={Sharafi, Zohreh and Sharif, Bonita and Gu{\'e}h{\'e}neuc, Yann-Ga{\"e}l and Begel, Andrew and Bednarik, Roman and Crosby, Martha},
  journal={Empirical Software Engineering},
  volume={25},
  number={5},
  pages={3128--3174},
  year={2020},
  publisher={Springer}
}

@inproceedings{burch2025eye,
  title={Eye Tracking Studies in Visualization: Phases, Guidelines, and Checklist},
  author={Burch, Michael and Kurzhals, Kuno and Weiskopf, Daniel},
  booktitle={Proceedings of the 2025 Symposium on Eye Tracking Research and Applications},
  pages={1--7},
  year={2025}
}

@article{grabinger2024eye,
  title={On eye tracking in software engineering},
  author={Grabinger, Lisa and Hauser, Florian and Wolff, Christian and Mottok, J{\"u}rgen},
  journal={SN Computer Science},
  volume={5},
  number={6},
  pages={729},
  year={2024},
  publisher={Springer}
}

@article{carter2020best,
  title={Best practices in eye tracking research},
  author={Carter, Benjamin T and Luke, Steven G},
  journal={International Journal of Psychophysiology},
  volume={155},
  pages={49--62},
  year={2020},
  publisher={Elsevier}
}

@article{monkaresi2016automated,
  title={Automated detection of engagement using video-based estimation of facial expressions and heart rate},
  author={Monkaresi, Hamed and Bosch, Nigel and Calvo, Rafael A and D'Mello, Sidney K},
  journal={IEEE Transactions on Affective Computing},
  volume={8},
  number={1},
  pages={15--28},
  year={2016},
  publisher={IEEE}
}

@INPROCEEDINGS{busjahn2015eye,
  author={Busjahn, Teresa and Bednarik, Roman and Begel, Andrew and Crosby, Martha and Paterson, James H. and Schulte, Carsten and Sharif, Bonita and Tamm, Sascha},
  booktitle={2015 IEEE 23rd International Conference on Program Comprehension}, 
  title={Eye Movements in Code Reading: Relaxing the Linear Order}, 
  year={2015},
  volume={},
  number={},
  pages={255-265},
  doi={10.1109/ICPC.2015.36}}

@INPROCEEDINGS{peitek2021program,
  author={Peitek, Norman and Apel, Sven and Parnin, Chris and Brechmann, André and Siegmund, Janet},
  booktitle={2021 IEEE/ACM 43rd International Conference on Software Engineering (ICSE)}, 
  title={Program Comprehension and Code Complexity Metrics: An fMRI Study}, 
  year={2021},
  volume={},
  number={},
  pages={524-536},
  doi={10.1109/ICSE43902.2021.00056}}

@INPROCEEDINGS{fritz2016leveraging,
  author={Fritz, Thomas and Müller, Sebastian C.},
  booktitle={2016 IEEE 23rd International Conference on Software Analysis, Evolution, and Reengineering (SANER)}, 
  title={Leveraging Biometric Data to Boost Software Developer Productivity}, 
  year={2016},
  volume={5},
  number={},
  pages={66-77},
  doi={10.1109/SANER.2016.107}}

@article{zhang2024eyetrans,
author = {Zhang, Yifan and Li, Jiliang and Karas, Zachary and Bansal, Aakash and Li, Toby Jia-Jun and McMillan, Collin and Leach, Kevin and Huang, Yu},
title = {EyeTrans: Merging Human and Machine Attention for Neural Code Summarization},
year = {2024},
issue_date = {July 2024},
publisher = {Association for Computing Machinery},
address = {New York, NY, USA},
volume = {1},
number = {FSE},
url = {https://doi.org/10.1145/3643732},
doi = {10.1145/3643732},
journal = {Proc. ACM Softw. Eng.},
month = jul,
articleno = {6},
numpages = {22}
}

@inproceedings{ko2009finding,
author = {Ko, Amy J. and Myers, Brad A.},
title = {Finding causes of program output with the Java Whyline},
year = {2009},
isbn = {9781605582467},
publisher = {Association for Computing Machinery},
address = {New York, NY, USA},
url = {https://doi.org/10.1145/1518701.1518942},
doi = {10.1145/1518701.1518942},
booktitle = {Proceedings of the SIGCHI Conference on Human Factors in Computing Systems},
pages = {1569–1578},
numpages = {10},
location = {Boston, MA, USA},
series = {CHI '09}
}

@INPROCEEDINGS{zyrianov2020automated,
  author={Zyrianov, Vlas and Guarnera, Drew T. and Peterson, Cole S. and Sharif, Bonita and Maletic, Jonathan I.},
  booktitle={2020 IEEE International Conference on Software Maintenance and Evolution (ICSME)}, 
  title={Automated Recording and Semantics-Aware Replaying of High-Speed Eye Tracking and Interaction Data to Support Cognitive Studies of Software Engineering Tasks}, 
  year={2020},
  volume={},
  number={},
  pages={464-475},
  doi={10.1109/ICSME46990.2020.00051}}

@article{bansal2023towards,
author = {Bansal, Aakash and Sharif, Bonita and McMillan, Collin},
title = {Towards Modeling Human Attention from Eye Movements for Neural Source Code Summarization},
year = {2023},
issue_date = {May 2023},
publisher = {Association for Computing Machinery},
address = {New York, NY, USA},
volume = {7},
number = {ETRA},
url = {https://doi.org/10.1145/3591136},
doi = {10.1145/3591136},
journal = {Proc. ACM Hum.-Comput. Interact.},
month = may,
articleno = {167},
numpages = {19}
}

@inproceedings{muller2016using,
author = {M\"{u}ller, Sebastian C. and Fritz, Thomas},
title = {Using (bio)metrics to predict code quality online},
year = {2016},
isbn = {9781450339001},
publisher = {Association for Computing Machinery},
address = {New York, NY, USA},
url = {https://doi.org/10.1145/2884781.2884803},
doi = {10.1145/2884781.2884803},
booktitle = {Proceedings of the 38th International Conference on Software Engineering},
pages = {452–463},
numpages = {12},
location = {Austin, Texas},
series = {ICSE '16}
}

@article{bednarik2012expertise,
  title={Expertise-dependent visual attention strategies develop over time during debugging with multiple code representations},
  author={Bednarik, Roman},
  journal={International Journal of Human-Computer Studies},
  volume={70},
  number={2},
  pages={143--155},
  year={2012},
  publisher={Elsevier}
}

@INPROCEEDINGS{paltenghi2021thinking,
  author={Paltenghi, Matteo and Pradel, Michael},
  booktitle={2021 36th IEEE/ACM International Conference on Automated Software Engineering (ASE)}, 
  title={Thinking Like a Developer? Comparing the Attention of Humans with Neural Models of Code}, 
  year={2021},
  volume={},
  number={},
  pages={867-879},
  doi={10.1109/ASE51524.2021.9678712}}

@article{deng2023eyettention,
  title={Eyettention: An attention-based dual-sequence model for predicting human scanpaths during reading},
  author={Deng, Shuwen and Reich, David R and Prasse, Paul and Haller, Patrick and Scheffer, Tobias and J{\"a}ger, Lena A},
  journal={Proceedings of the ACM on Human-Computer Interaction},
  volume={7},
  number={ETRA},
  pages={1--24},
  year={2023},
  publisher={ACM New York, NY, USA}
}

@article{su2026code,
  title={Do code llms do static analysis?},
  author={Su, Chia-Yi and McMillan, Collin},
  journal={Empirical Software Engineering},
  volume={31},
  number={5},
  pages={116},
  year={2026},
  publisher={Springer}
}

@article{ding2024code,
  title={Do code summarization models process too much information? function signature may be all that is needed},
  author={Ding, Xi and Peng, Rui and Chen, Xiangping and Huang, Yuan and Bian, Jing and Zheng, Zibin},
  journal={ACM Transactions on Software Engineering and Methodology},
  volume={33},
  number={6},
  pages={1--35},
  year={2024},
  publisher={ACM New York, NY}
}

@book{wohlin2012experimentation,
  title={Experimentation in software engineering},
  author={Wohlin, Claes and Runeson, Per and H{\"o}st, Martin and Ohlsson, Magnus C and Regnell, Bj{\"o}rn and Wessl{\'e}n, Anders},
  year={2012},
  publisher={Springer Science \& Business Media}
}

@article{engbert2005swift,
  title={SWIFT: a dynamical model of saccade generation during reading.},
  author={Engbert, Ralf and Nuthmann, Antje and Richter, Eike M and Kliegl, Reinhold},
  journal={Psychological review},
  volume={112},
  number={4},
  pages={777},
  year={2005},
  publisher={American Psychological Association}
}

@inproceedings{nilsson2009learning,
  title={Learning where to look: Modeling eye movements in reading},
  author={Nilsson, Mattias and Nivre, Joakim},
  booktitle={Proceedings of the Thirteenth Conference on Computational Natural Language Learning (CoNLL-2009)},
  pages={93--101},
  year={2009}
}

@inproceedings{nilsson2011entropy,
  title={Entropy-driven evaluation of models of eye movement control in reading},
  author={Nilsson, Mattias and Nivre, Joakim},
  booktitle={Proceedings of the 8th International NLPCS Workshop},
  pages={201--212},
  year={2011}
}

@article{reichle1998toward,
  title={Toward a model of eye movement control in reading.},
  author={Reichle, Erik D and Pollatsek, Alexander and Fisher, Donald L and Rayner, Keith},
  journal={Psychological review},
  volume={105},
  number={1},
  pages={125},
  year={1998},
  publisher={American Psychological Association}
}

\newpage
\begin{comment}
    \section{Biography Section}
If you have an EPS/PDF photo (graphicx package needed), extra braces are
 needed around the contents of the optional argument to biography to prevent
 the LaTeX parser from getting confused when it sees the complicated
 $\backslash${\tt{includegraphics}} command within an optional argument. (You can create
 your own custom macro containing the $\backslash${\tt{includegraphics}} command to make things
 simpler here.)
 
\vspace{11pt}

\bf{If you include a photo:}\vspace{-33pt}
\begin{IEEEbiography}[{\includegraphics[width=1in,height=1.25in,clip,keepaspectratio]{fig1}}]{Michael Shell}
Use $\backslash${\tt{begin\{IEEEbiography\}}} and then for the 1st argument use $\backslash${\tt{includegraphics}} to declare and link the author photo.
Use the author name as the 3rd argument followed by the biography text.
\end{IEEEbiography}

\vspace{11pt}

\bf{If you will not include a photo:}\vspace{-33pt}
\begin{IEEEbiographynophoto}{John Doe}
Use $\backslash${\tt{begin\{IEEEbiographynophoto\}}} and the author name as the argument followed by the biography text.
\end{IEEEbiographynophoto}
\end{comment}

\vfill

\end{document}